\documentclass[11pt]{article}
\pdfoutput = 1

\usepackage{jheppub} 

\usepackage[T1]{fontenc} 
\usepackage{color,graphicx}
\usepackage{amsmath}
\usepackage{amssymb}
\usepackage{verbatim}
\usepackage{enumerate}
\usepackage{stmaryrd}
\usepackage{physics}
\usepackage{amsfonts}
\usepackage{array}
\usepackage{setspace}
\usepackage{float}
\usepackage{url}
\usepackage{mathtools}
\usepackage{bbold}
\usepackage{tikz}
\usepackage{wrapfig}
\usepackage{subfigure}
\usepackage[labelfont=bf,font={rm},font={small}]{caption}
\usepackage{cancel}
\usepackage{nicematrix}
\usepackage{xcolor}
\allowdisplaybreaks[4]
\usepackage[most]{tcolorbox}
\usepackage{braket}
\usepackage{todonotes}
\usepackage{appendix}
\usepackage{comment}

\newcommand{\be}{\begin{equation}}
\newcommand{\ee}{\end{equation}}
\newcommand{\bea}{\begin{align}}
\newcommand{\eea}{\end{align}}
\newcommand{\bi}{\begin{itemize}}
\newcommand{\ei}{\end{itemize}}

\newcommand{\Ram}{{\mathfrak R}}
\newcommand{\cS}{{\cal S}}

\newcommand{\cL}{{\cal L}}

\numberwithin{equation}{section}

\title{\boldmath $q$-deformed Weil--Petersson volumes from the spectral edge of DSSYK and supersymmetric extensions}

\author[a]{Wasif Ahmed}
\author[b,c]{Joydeep Naskar}
\author[a]{Krishan Saraswat}

\affiliation[a]{Physics Department, Broida Hall, University of California, Santa Barbara, CA 93106, USA}
\affiliation[b]{Beijing Institute of Mathematical Sciences and Applications, Huaibei Town, Huairou District, Beijing 101408, China}
\affiliation[c]{Kavli Institute for Theoretical Physics, Santa Barbara, CA 93106, USA}

\emailAdd{wasif@ucsb.edu}
\emailAdd{joydeepnaskar@bimsa.cn}
\emailAdd{ksaraswa@ualberta.ca}

\abstract{The $q$-deformed Weil--Petersson volumes of Do and Norbury are the top-degree parts of Okuyama's discrete volumes, which are defined by topological recursion on the spectral curve of the double-scaled SYK (DSSYK) model. We show that the same $q$-deformed volumes can also be obtained from a double-scaled matrix model whose disk density is found by zooming in on the soft edges of the DSSYK density. Applying the same procedure to the hard edge of $\mathcal{N}=1$ DSSYK, we obtain candidate $q$-deformed $\mathcal{N}=1$ super Weil--Petersson volumes. They are polynomials of fixed degree, and they reduce to $\mathcal{N}=1$ super Weil--Petersson volumes in a scaled $q\to1$ limit. These candidate volumes differ from the $\mathcal{N}=1$ volumes of Do and Norbury in~\cite{DoNorbury:2025qMirzakhani} up to what we call ``irrelevant'' contributions which exactly vanish in the scaled $q\to 1$ limit. Together, these results demonstrate how $q$-deformed volumes can be naturally associated to the spectral edges of various DSSYK models and provides an intuitive way to construct them in supersymmetric contexts.}

\begin{document}

\maketitle

\section{Introduction}
\label{sec:introduction}

\subsection{Background and context}
In the past decade 2D dilaton theories of gravity, such as Jackiw--Teitelboim (JT) gravity~\cite{Teitelboim:1983ux,Jackiw:1984je,Grumiller:2002nm,Grumiller:2021cwg,Mertens:2022irh}, have played a pivotal role in many developments in quantum gravity ranging from their appearance in the near horizon dynamics of near extremal black holes to their uses as a testing ground to better understand the gravitational path integral~\cite{Almheiri:2014cka,Maldacena:2016upp,Cotler:2016fpe,Stanford:2017thb,Almheiri:2019qdq,Iliesiu:2020qvm,Heydeman:2020hhw,Blommaert:2021fob,Saraswat:2021ong,Boruch:2022tno,Iliesiu:2022kny,Boruch:2023trc,Cassani:2025sim,Saraswat:2026cuo}. In particular, the realization that the Euclidean path integral of JT gravity has a double scaled matrix integral formulation has led to a surge of interest in Weil--Petersson (WP) volumes, denoted $V^{WP}_{g,n}(b_1,..,b_n)$, which enable the efficient computation of various higher genus corrections of quantities derived from the path integral in terms of trumpets glued to bordered hyperbolic Riemann surfaces of genus-$g$ with $n$ geodesic boundaries of lengths $b_i$~\cite{Saad:2019lba,Stanford:2019vob,Turiaci:2023jfa}. Weil--Petersson volumes themselves have a rich history in mathematical literature~\cite{Wolpert85,Wolpert1986,Penner92,Zograf1993TheWV,zograf1998weilpeterssonvolumesmodulispaces,manin1999invertiblecohomologicalfieldtheories} and their explicit computation by Mirzakhani showed that the volumes are polynomials in the squared boundary lengths and derived a recursion relation that determines them~\cite{Mirzakhani:2006simple,Mirzakhani:2007intersection}. Subsequently, these volumes have also been shown to fit into the framework of topological recursion going back to Chekhov--Eynard--Orantin recursion~\cite{Chekhov:2006vd,Eynard:2007kz,EynardOrantin:2007WP}. The key input (initial condition) is the specification of a spectral curve in the complex plane (parameterized by $z$) which is given by\footnote{Technically we also need to specify $\omega_{0,2}(z_1,z_2)$. Throughout this entire paper we will exclusively only deal with $\omega_{0,2}(z_1,z_2)=\frac{dz_1dz_2}{(z_1-z_2)^2}$ and only $x(z)$ and $y(z)$ will change depending on the specific model we are discussing.}:
\begin{equation}
\label{JTGravitySpecCurve}
    x^{WP}(z)=\frac{1}{2}z^2\ ,\qquad y^{WP}(z)=\frac{\sin(2\pi z)}{2\pi}\ .
\end{equation}
The machinery of topological recursion uses this curve to produce differential $n$-forms $\omega^{WP}_{g,n}(z_1,..,z_n)$ whose inverse Laplace transform from $(z_1,...,z_n)\to (b_1,..,b_n)$ reproduces the Weil--Petersson volumes, $V^{WP}_{g,n}(b_1,..,b_n)$\footnote{For the convenience of the reader we provide an essential overview of how topological recursion is used for a given spectral curve in Appendix \ref{sec:topological-recursion} based on lecture notes~\cite{Bouchard:2024fih}.}.   

Alternatively, there is another more recently developed formalism which can also be used to reproduce Weil--Petersson volumes~\cite{Johnson:2024vms,Johnson:2024svms,Johnson:2025furtherSVMS,Ahmed:2025lxe,Johnson:2026jbq,Johnson:2026twg,Johnson:2026jls,Johnson:2026jgs,Johnson:2026wlo}. The origins of this approach can be traced back to the study of random matrix models through the introduction of orthogonal polynomials which enable the evaluation of various matrix integrals in terms of orthogonal polynomial recursion coefficients~\cite{mehta2004random,Livan_2018,Van_Assche_2020}. When double scaled, the orthogonal polynomials are repackaged into wavefunctions which appear in an auxiliary Schrodinger problem~\cite{Brezin:1990rb,Douglas:1989ve,Gross:1989vs,Ginsparg:1993is,Johnson:2020npjt,Johnson:2020mwi,Johnson:2021jtsugra,JohnsonRossoSvesko:2021doublecut,Johnson:2021rsh,Johnson:2022wsr,Johnson:2021explore,JohnsonUsatyuk:2025gaps,Johnson:2024n2,Johnson:2024vms,Johnson:2024svms,Johnson:2025furtherSVMS,Johnson:2025oty,Ahmed:2025lxe}:
\begin{equation}
\label{SchrodingerEq}
    \mathcal{H}\psi(E,x)=E\psi(E,x)\ , \qquad \mathcal{H}=-\hbar^2\frac{\partial^2}{\partial x^2}+u(x)\ .
\end{equation}
The auxiliary system is defined in the range $x\in(-\infty,\mu)$ where $\mu$ is referred to as the Fermi surface. The potential, $u(x)$, is governed by the so called string equation given as~\cite{Morris:1990cq,DalleyJohnsonMorris:1992mcm,DalleyJohnsonMorrisWatterstam:1992umm}:    
\begin{equation}
    \begin{split}
    \label{FullStringEquation}
        &u\mathcal{R}-\frac{\hbar^2}{2}\mathcal{R}\mathcal{R}''+\frac{\hbar^2}{4}\mathcal{R}'^2=\hbar^2\Gamma^2\ , \\
        &\mathcal{R}=\sum_{k=1}^\infty t_k R_k[u]+x\ ,\\
    \end{split}
\end{equation}
where primes are derivatives with respect to $x$ and powers of $\hbar$ should be thought of as a genus counting parameter in the perturbative genus expansion, $R_k[u]$ are Gel’fand Dikii polynomials\footnote{For a more modern and explicit overview of these polynomials and their constructions the reader can
look at appendices of~\cite{Ahmed:2025lxe}.}~\cite{Gelfand:1975rn,GelfandDikii1977,Gelfand:1995qu}, and $t_k$ are called KdV times which are fixed by using the disk (genus 0) matrix model spectral density~\cite{Johnson:2020npjt,Johnson:2022wsr,Johnson:2025oty}, denoted $\rho_0(E)$\footnote{See Appendix \ref{sec:ODE-method} for a quick review of the details to fix $t_k$ using some disk density of state, $\rho_0(E)$.}. For the purpose of reproducing Weil--Petersson volumes the relevant disk density is that of 2D JT gravity given by:
\begin{equation}
\label{JTGravityDOS}
    \rho_0(E)=\frac{\sinh(2\pi\sqrt{E})}{4\pi^2\hbar}\ .
\end{equation}
Using this density of state one can fix the KdV times ($t_k$) and write an equation for $u_0(x)$\footnote{Where $u_0(x)$ appears as the leading term (genus 0) in the perturbative expansion of $u(x)=u_0+\sum_{g=1}^\infty\hbar^{2g}u_{2g}(x)$.}. It turns out that the Weil-Peterson volumes can then be expressed entirely in terms of the derivatives of $u_0$ at the Fermi surface $x=\mu$. This alternative formalism to arrive at the Weil-Peterson volumes will be referred to as ``the ODE method'' to distinguish it from the formalism of topological recursion. 

Although the ODE method and topological recursion are distinct approaches to obtain Weil--Petersson volumes their starting points are intimately related to each other. In particular, the spectral curve $(y(z),x(z))$ can be used to construct the density used in the ODE approach through a map of the form $\rho_0(E)=\frac{\eta}{\pi\hbar} y(z(x))\vert_{x=\gamma E}$ where $\eta$ and $\gamma$ are fixed through a matching procedure outlined in Appendix \ref{sec:matching-ODE-TR}. For example, the JT gravity density of states in Eq. (\ref{JTGravityDOS}) can be constructed from the spectral curve in Eq. (\ref{JTGravitySpecCurve}) by using $\gamma=-\frac{1}{2}$ and $\eta=-\frac{i}{2}$ (which reproduces the commonly used $z=-E^2$ relation~\cite{Saad:2019lba,Johnson:2024vms,Johnson:2024svms,Ahmed:2025lxe,Johnson:2026twg,Johnson:2026jls} along with $y(z)=2\pi i\hbar \rho_0(E=-z^2)$~\cite{Eynard:2014zxa}). In addition, both approaches have also been valuable in the study and computation of supersymmetric extensions of Weil--Petersson volumes (or super Weil--Petersson volumes) which arise from the study of the supermoduli space of super Riemann surfaces as well as physical considerations of supersymmetric analogues of JT gravity~\cite{Norbury:2020vyi,Stanford:2019vob,Turiaci:2023jfa}. 

In recent years, there has been increased interest in generalizing the notion of Weil--Petersson volumes to more discrete settings~\cite{Norbury:2008lattice,NorburyScott:2013polynomials,Okuyama:2023discrete,Giacchetto:2025nnb}. In particular, discussions of discrete analogues of Weil--Petersson volumes can be traced back to the mathematical constructions of Norbury and Scott~\cite{Norbury:2008lattice,NorburyScott:2013polynomials}; and more recently Okuyama was able to arrive at the same results from a more physically motivated construction involving the double scaled SYK (DSSYK) model~\cite{Okuyama:2023discrete}. Essentially, the DSSYK model can be associated to a certain kind of random matrix model whose spectral curve is given by~\cite{Okuyama:2023discrete,DoNorbury:2025dssyk,Jafferis:2022JTmatter}:
\begin{eqnarray}
\label{eq:Do-Norbury-full-curve}
    x(z)&=&z+z^{-1}\ ,\nonumber\\
    y(z)&=&\frac{1}{2}(z-z^{-1})
    \prod_{m=1}^{\infty}
    \frac{(1-q^m z^2)(1-q^m z^{-2})}{(1-q^m)^2}\ ,
\end{eqnarray}
where $0<q<1$ (a parameter that is arises in the double scaling limit of the SYK model). Inputting this spectral curve into the framework of topological recursion yields the differential $n$-forms, $\omega^q_{g,n}(z_1,..,z_n)$, which can then be expanded around $z_i=0$ to yield an expansion of the form:
\begin{equation}
\label{DiscreteWPVolumesDefFromomega}
\omega^q_{g,n}(z_1,..,z_n)=\sum_{b_1,..,b_n=1}^\infty N^q_{g,n}(b_1,..,b_n)\prod_{i=1}^n b_i z_i^{b_i-1}dz_i\ ,
\end{equation}
where $N^q_{g,n}(b_1,..,b_n)$ are referred to as discrete analogues of Weil--Petersson volumes\footnote{In recent work~\cite{Giacchetto:2025nnb}, it was shown that notions of ``discrete'' volumes of the moduli space of Riemann surfaces appears quite generally from the pruned correlators of one-matrix models. The matrix model dual of the DSSYK model is one example of this which the authors explored to prove a number of Okuyama's conjectures~\cite{Okuyama:2023discrete}.}. Unlike the ordinary Weil--Petersson volumes, which were polynomials in the boundary $b_i$, the discrete analogues are quasi-polynomials in discrete boundary lengths $b_i\in\mathbb{N}$, with parity dependent coefficients involving the $q$-analogue of the zeta function defined by the following series:
\begin{equation}
\label{qanalogueZetaDef}
    \zeta_q(2m)=\sum_{n=1}^\infty \frac{q^{nm}}{(1-q^n)^{2m}}\ ,
\end{equation}
which satisfies the following $q\to 1$ limit:
\begin{equation}
    \lim_{q\to 1}(1-q)^{2m}\zeta_q(2m)=\zeta(2m)\ .
\end{equation}
Using this, Okuyama demonstrated that in a scaled limit as $q\to 1$ the discrete volumes he defined will reproduce the familiar Weil--Petersson volumes:
\begin{equation}
\label{ScalinglimitqDefWPVols}
    \lim_{q\to 1}(1-q)^{2(3g-3+n)}N^q_{g,n}(b_{1,q},..,b_{n,q})=V^{WP}_{g,n}(b_1,..,b_n)\ ,\qquad b_{i,q}=\frac{b_{i}}{1-q}\ .
\end{equation}
In subsequent work, Do and Norbury introduced the grading:
\begin{equation}
\label{eq:q-degree-grading}
    \deg b_i=1\ ,\qquad \deg\zeta_q(2k)=2k\ ,
\end{equation}
and rigorously proved $N^q_{g,n}(b_1,..,b_n)$ has degree $6g{-}6{+}2n$ for every stable pair $(g,n)$~\cite{DoNorbury:2025dssyk}. A particularly important observation is that only the highest degree terms play a non-trivial role in reproducing the Weil-Peterson volumes the lower degree parts evaluate to zero in the scaled $q\to 1$ limit. It is precisely these highest degree terms of $N^q_{g,n}(b_1,..,b_n)$, which we denote as $V^{qWP}_{g,n}(b_1,..,b_n)$, which are called q-deformed Weil--Petersson volumes, which will be the central focus of this work. Do and Norbury also demonstrated that these q-deformed Weil--Petersson volumes can be constructed within the framework of topological recursion using the spectral curve~\cite{DoNorbury:2025dssyk}: 
\begin{equation}
\label{qdefSpectralCuve}
    x^{qWP}(z)=\frac{1}{2}z^2\ , \qquad y^{qWP}(z)=z\exp\left[-\sum_{m=1}^\infty \frac{\zeta_q(2m)}{m}(4z^2)^m\right]\ .
\end{equation}
 They also wrote q-analogues of Mirzakhani’s recursion for Weil–Petersson volumes as well as recursion relations to for q-deformed $\mathcal{N}=1$ supersymmetric contexts~\cite{DoNorbury:2025qMirzakhani}.

\subsection{Goals and results of this work}
What is clear from the review done in the previous subsection is that expressions for q-deformed Weil--Petersson volumes appear as the highest degree terms of the discrete analogues of Weil--Petersson volumes, which themselves can be obtained using the spectral curve of the DSSYK model~\cite{Okuyama:2023discrete,DoNorbury:2025dssyk}. Furthermore, the work of Do and Norbury~\cite{DoNorbury:2025dssyk} and their identification of the spectral curve in Eq. (\ref{qdefSpectralCuve}) that isolates these highest degree terms demonstrates that q-deformed volumes can be defined on their own without necessarily appealing to a discrete analogue of volumes. Indeed, it seems that Do and Norbury adopt this view in when defining the $\mathcal{N}=1$ extension of q-deformed Weil--Petersson volumes (they did not use any kind of SYK starting point to define them)~\cite{DoNorbury:2025qMirzakhani}. This leads to an interesting series of questions which we explore in this work. 

One question pertains to the physical interpretation of q-deformed Weil--Petersson from the perspective of the DSSYK model. It is clear that the discrete Weil--Petersson volumes arise from the matrix model dual of the DSSYK model, however, is there also a similar interpretation of for the q-deformed Weil--Petersson volumes (in particular, we are interested in an interpretation that does not need to appeal to any discrete analogue of Weil--Petersson volumes). In Section \ref{qWPVolFromEdgeDSSYKSection}, we will demonstrate that the answer to the question is yes, the q-deformed Weil--Petersson volumes arise from a double scaled matrix model with a leading disk density of states given by\footnote{The terminology can be slightly confusing in this discussion since the term ``double scaling'' is used in slightly different ways. To clarify, the ``double scaled'' terminology in DSSYK has to do with the how two parameters are SYK model are simultaneously scaled to defined the DSSYK model, the matrix model which is dual to the DSSYK is not a double scaled matrix model. However, we can take a double scaled limit near the edges of the matrix model which is dual to DSSYK to obtain a proper double scaled matrix model with a density given by Eq. (\ref{qDefVolDOS}).}:
\begin{equation}
\label{qDefVolDOS}
    \rho_0(E)=\frac{1}{2\pi\hbar}\sqrt{E}\,
    \exp\left[-\sum_{m=1}^{\infty}
    \frac{(-4)^m\zeta_q(2m)}{m}E^m\right]\ .
\end{equation}
We will explicitly demonstrate that using this density of states in the ODE formalism reproduces q-deformed Weil--Petersson volumes. Furthermore, we will demonstrate that the spectral density in Eq. (\ref{qDefVolDOS}) should be thought of as living on the soft edges of the DSSYK model. Based on this physical picture we will provide a prescription by which to go from the full DSSYK model density of states to the density of states we write in Eq. (\ref{qDefVolDOS}). 

Building on these physical insights we then pose questions pertaining to the construction of $\mathcal{N}=1$ q-deformed volumes of Do and Norbury~\cite{DoNorbury:2025qMirzakhani}. In particular, we can ask if the q-deformed $\mathcal{N}=1$ volumes of Do and Norbury can also be regarded as coming from a double scaled random matrix model that lives on the spectral edge of the $\mathcal{N}=1$ DSSYK model. We explore this in Section \ref{N=1DSSYKWPVolSection} of the paper. We begin by considering the hard edge of the $\mathcal{N}=1$ DSSYK model which naturally defines a double scaled random matrix model with a density given by:
\begin{equation}
\label{OurDensityForN=1qDefVols}
    \rho_0(E)=\frac{1}{2\pi\hbar\sqrt{E}}\exp\left[-\sum_{m=1}^\infty \frac{4^m\xi_q(2m)}{m}E^m  \right]\ .
\end{equation}
Where the q-deformation function $\xi_q$ can be written as the following series:
\begin{equation}
\label{xiDefintion}
    \xi_q(2m)=\sum_{n=1}^\infty\frac{(-q)^{nm}}{(1+(-q)^n)^{2m}}\ .
\end{equation}
When we apply the ODE method to generate expressions for the associated volumes, denoted $V^{qSWP}_{g,n}(b_1,..,b_n)$, we find that they are different from the ones that one would obtain through Do and Norbury's construction in~\cite{DoNorbury:2025dssyk}. Nonetheless, the volumes one obtains from the spectral density given in Eq. (\ref{OurDensityForN=1qDefVols}) will still satisfy the key features of the volumes given by Do and Norbury's construction. Namely, that each term in $V^{qSWP}_{g,n}(b_1,..,b_n)$ has a fixed degree of $2g-2$\footnote{Where we take the grading $\text{deg}(\xi_q(2m))=2m$ and $\text{deg}(b_i)=1$.}. Most importantly, the following scaled limit as $q\to 1$ reproduces the standard $\mathcal{N}=1$ super Weil--Petersson volumes:
\begin{equation}
    \lim_{q\to 1}(1-q)^{2g-2}V^{qSWP}_{g,n}(b_{1,q},..,b_{n,q})=V_{g,n}^{SWP}(b_1,..,b_n)\ , \qquad b_{i,q}=\frac{b_i}{1-q}\ .
\end{equation}
We demonstrate that the relation of these volumes to the ones of Do and Norbury is given by a simple replacement of $\xi_q(2m)\to (-1)^m\zeta^{odd}_q(2m)$ which can be understood by decomposing the series representation of $\xi_q$ into even and odd sectors and demonstrating the irrelevance (i.e. vanishing) of the even sector in the $q\to 1$ scaling limit. This leads to the conclusion that the Do and Norbury construction of $\mathcal{N}=1$ supersymmetric q-deformed volumes can also be obtained using the ODE method with the following a double scaled matrix model density of states:
\begin{equation}
\label{DoNorburyVolumesDOS}
    \rho^{DN}_0(E)=\frac{1}{2\pi\hbar\sqrt{E}}\exp\left[-\sum_{m=1}^\infty \frac{(-4)^m\zeta_q^{odd}(2m)}{m}E^m\right]\ .
\end{equation}
Which further demonstrates that this double scaled matrix model can be physically understood as living near the hard edge of the $\mathcal{N}=1$ DSSYK model up to a subtle discarding of the ``irrelevant parts'' of $\xi_q$ (i.e. discarding the even sector and leaving the essential $\zeta_q^{odd}$ of the Do and Norbury construction). We conclude Section \ref{N=1DSSYKWPVolSection} by posing the question of whether the $\mathcal{N}=1$ q-deformed volumes admit to an interpretation of being the highest degree terms of some discrete analogue of super Weil--Petersson volumes. The analysis presented in Section \ref{N=1DiscreteWPVolSec} suggests that spectral curve one obtains from analyzing the $\mathcal{N}=1$ DSSYK model contains a problematic structure which generates terms in $\omega_{g,n}$ that have an ill defined degree thereby preventing a construction of discrete volumes with properties resembling those of Okuyama's discrete volumes in the non-supersymmetric case. 

We conclude the work in Section \ref{DiscussionConclusionSection} by proposing that soft edges in the DSSYK model should naturally lead to a notion of q-deformed Weil--Petersson volumes and hard edges should lead to a notion of $\mathcal{N}=1$ q-deformed super Weil--Petersson volumes and that considering other supersymmetric DSSYK models such as $\mathcal{N}=2$ DSSYK should also lead to some natural definition of q-deformed $\mathcal{N}=2$ super Weil--Petersson volumes. We also pose further questions and possible avenues of investigation. 

\section{q-deformed Weil--Petersson volumes from the edge of DSSYK}
\label{qWPVolFromEdgeDSSYKSection}
In this section, we will discuss the construction of the spectral density of a double scaled matrix model that will reproduce q-deformed Weil--Petersson volumes using the ODE method. Our starting point will be the same as Okuyama's, namely the SYK model~\cite{Maldacena:2016hyu,Garcia-Garcia:2016mno,Polchinski:2016xgd,Sarosi:2017ykf} whose Hamiltonian is given as: 
\begin{equation}
    H=i^{p/2}\sum_{1\leq i_1<\cdot\cdot\cdot<i_p \leq N} J_{i_1\cdot\cdot\cdot i_p}\psi_{i_1}\cdot\cdot\cdot\psi_{i_p}\ ,
\end{equation}
where $N\in 2\mathbb{N}$ is the number of Majorana fermions (which satisfy $\{\psi_{i},\psi_j\}=2\delta_{i,j}$), $p\leq N$ is the number of fermions that randomly interact via $J_{i_1,..,i_p}$ which itself is a Gaussian random variable (completely anti-symmetric under the exchange of indices) with :
\begin{equation}
    \braket{J_{i_1\cdot\cdot\cdot i_p}}_J=0\ , \qquad \braket{J_{i_1\cdot\cdot\cdot i_p}^2}_J=\binom{N}{p}^{-1}\ .
\end{equation}
The so called double scaled SYK (DSSYK) model~\cite{Berkooz:2018qkz,Berkooz:2018jqr,Berkooz:2020xne,Lin:2022rbf,Lin:2023trc,Berkooz:2024lgq} studies the model in the limit given below:
\begin{equation}
    N,p\to\infty\ ,\qquad \text{with}\qquad \lambda=\frac{2p^2}{N}=\text{fixed}\ . 
\end{equation}
In this limit, ensemble averages of the $k$-th moment of the Hamiltonian are given by:
\begin{equation}
    \braket{\text{Tr}(H^k)}_J=\sum_{\text{chord diagrams}} q^{\text{\# intersections}}\ ,\qquad q=e^{-\lambda}\ .
\end{equation}
Using the moments of the Hamiltonian one can construct an expression for the ensemble averaged canonical partition function, $\braket{\text{Tr}e^{-\beta H}}_J$, which is given by the following integral:
\begin{equation}
\begin{split}
    &\braket{\text{Tr}e^{-\beta H}}_J=\int_0^\pi \frac{d\theta}{2\pi}\left(q,e^{\pm2i\theta};q\right)_\infty e^{-\beta E(\theta)}\ ,\\
    &E(\theta)=-a\cos\theta\ ,\qquad a=\frac{2}{\sqrt{1-q}}\ ,\\
\end{split}
\end{equation}
where one defines:
\begin{equation}
\begin{split}
    &\left(q,e^{\pm 2 i \theta};q\right)_{\infty}=(q;q)_\infty(e^{2i\theta},q)_\infty(e^{-2i\theta};q)_\infty\ ,\\
    &(a,q)_\infty= \prod_{k=0}^\infty (1-aq^k)\ .\\
\end{split}
\end{equation}
Changing integration variables of the integral from $\theta$ to $E$ allows one to read off the density of states of the DSSYK model to be:
\begin{equation}
\label{DSSYKFullDOS}
    \rho_0(E)=\frac{\left(q,e^{\pm 2 i\theta(E)};q\right)_\infty}{2\pi a\sin\theta(E)}=\frac{2 (q;q)_\infty^3}{\pi a^2}\sqrt{a^2-E^2}\prod_{n=1}^\infty \left[1-\frac{q^n}{(1-q^n)^2}\frac{4(a^2-E^2)}{a^2}\right]\ .
\end{equation}
This was taken as the disk density of states for a dual random matrix model description~\cite{Okuyama:2023discrete,Jafferis:2022JTmatter} and this was used to define Okuyama's spectral curve in Eq. (\ref{eq:Do-Norbury-full-curve}). Rather than taking this step we will stop here and rewrite this spectral density as follows:
\begin{equation}
    \begin{split}
        &\rho_0(E)=\mathcal{C}\sqrt{a^2-E^2}\exp\left[\ln\left(\prod_{n=1}^\infty \left[1-\frac{q^n}{(1-q^n)^2}\frac{4(a^2-E^2)}{a^2}\right]\right)\right]\\
        &=\mathcal{C}\sqrt{a^2-E^2}\exp\left[\sum_{n=1}^\infty \ln\left(1-\frac{q^n}{(1-q^n)^2}\frac{4(a^2-E^2)}{a^2}\right)\right]\\
        &=\mathcal{C}\sqrt{a^2-E^2}\exp\left[-\sum_{n=1}^\infty \sum_{m=1}^\infty \frac{\frac{q^{nm}}{(1-q^n)^{2m}}}{m}\left(\frac{4(a^2-E^2)}{a^2}\right)^m\right]\\
        &=\mathcal{C}\sqrt{a^2-E^2}\exp\left[-\sum_{m=1}^\infty\frac{\zeta_q(2m)}{m}\left(\frac{4(a^2-E^2)}{a^2}\right)^m\right]\ ,\\
    \end{split}
\end{equation}
where we simply defined $\mathcal{C}$ as the q-dependent pre-factor appearing in front of the density. Going from the second to third line involves using the series expansion of $\ln(1-x)=-\sum_{m=1}^\infty \frac{x^m}{m}$ and going from the third line to the fourth line involves the interchange of the series in $m$ and $n$ and using the definition of the q-analogue of the zeta function reviewed in Eq. (\ref{qanalogueZetaDef}). This particular representation of the DSSYK density of states makes manifest the appearance of $\zeta_q(2m)$. Furthermore, it is worth observing the the density of states that arises from DSSYK model is supported on the finite interval $E\in[-a,a]$ with two soft edges. In particular, this density is not characteristic of a matrix model which is double scaled. However, it provides the perfect starting point to construct one using the soft edges located at $E=\pm a$. To be concrete let us focus on the lower soft edge at $E=-a$. We define $E(\epsilon)=-a+\epsilon$. Then a leading truncation of the DSSYK density for small $\epsilon$ gives density with the following structure:
\begin{equation}
    \rho_0(\epsilon)=\mathcal{C}'\sqrt{\epsilon}\exp\left[-\sum_{m=1}^\infty \frac{\zeta_q(2m)}{m}(\gamma\epsilon)^m\right]\ ,
\end{equation}
where $\mathcal{C}'$ and $\gamma$ are some constants which will be fixed shortly. The structure of this density which we obtained by zooming in on the soft edge of the DSSYK model describes the ``essence'' of a double scaled random matrix model which will generate the q-deformed Weil--Petersson volumes. To understand this, one can consider the following scaling limit of the edge density we obtained:
\begin{equation}
    \begin{split}
        &\lim_{q\to 1}\frac{\rho_0((1-q)^2\epsilon)}{1-q}=\mathcal{C}'\sqrt{\epsilon}\lim_{q\to 1}\exp\left[-\sum_{m=1}^\infty \frac{(1-q)^{2m}\zeta_q(2m)}{m}\left(\gamma\epsilon\right)^m \right]\\
        &=\mathcal{C'}\sqrt{\epsilon}\exp\left[-\sum_{m=1}^\infty \frac{\zeta(2m)}{m}\left(\gamma\epsilon\right)^m\right]\\
        &=\mathcal{C'}\sqrt{\epsilon}\exp\left[-\sum_{m=1}^\infty \frac{\zeta(2m)}{m}\left(\gamma\epsilon\right)^m\right]\\
        &=\frac{\mathcal{C}'\sin\left(\pi\sqrt{\gamma\epsilon}\right)}{\pi\sqrt{\gamma}}\ .\\
    \end{split}
\end{equation}
We can reproduce the disk density of state of JT gravity given in Eq. (\ref{JTGravityDOS}) by fixing $\gamma=-4$ and $\mathcal{C}'=(2\pi\hbar)^{-1}$. Which gives us the promised density of state in Eq. (\ref{qDefVolDOS}):
\begin{equation}
    \rho_0(E)=\frac{1}{2\pi\hbar}\sqrt{E}\,
    \exp\left[-\sum_{m=1}^{\infty}
    \frac{(-4)^m\zeta_q(2m)}{m}E^m\right]\ .
\end{equation}
Since we already know, by construction, this density will reproduce the JT gravity density of states in a scaled $q\to 1$ limit we also should expect the ODE method with this particular density to reproduce the correct q-deformed volumes whose scaled $q\to 1$ limit gives the Weil--Petersson volumes. In particular, in Appendix \ref{sec:top-first-volumes} we explicitly demonstrate this in a number of sample calculations. This approach to obtain q-deformed Weil--Petersson volumes does not need to appeal to any discrete analogue of Weil--Petersson volumes and highlights how one can physically view q-deformed Weil--Petersson volumes as objects associated to the local structure of the soft spectral edge of the DSSYK model. This is in contrast to the discrete analogues of Weil--Petersson volumes which actually requires global knowledge of the density to define Okuyama's spectral curve. 

To conclude this section, we will mention that the work of Do and Norbury~\cite{DoNorbury:2025dssyk} also corroborates this story since the procedure to obtain the spectral curve in Eq. (\ref{qdefSpectralCuve}) involved an expansion around the ramification points of Okuyama's spectral curve which in the language of the associated spectral density is really an expansion around the soft edges. This can demonstrated further by using the spectral curve given by Do and Norbury and then applying the matching procedure outlined in Appendix \ref{sec:matching-ODE-TR} to exactly reproduce the density in Eq. (\ref{qDefVolDOS}).   

\section{q-deformed super Weil--Petersson volumes from the edge of $\mathcal{N}=1$ DSSYK}
\label{N=1DSSYKWPVolSection}
In this section, we will extend our discussion of q-deformed volumes from the edge DSSYK models to a supersymmetric setting using the $\mathcal{N}=1$ DSSYK model as a starting point of our construction. 
\subsection{Global structure of $\mathcal{N}=1$ DSSYK density of states}
The starting point of our discussion is the $\mathcal{N}=1$ analogue of Eq. (\ref{DSSYKFullDOS}) which we quote from the work~\cite{Berkooz:2020xne}:
\begin{equation}
    \begin{split}
        &\rho_0(E)=\frac{(-q,e^{\pm 2i\theta};-q)_\infty}{\pi a \sin(2\theta)}\ ,\\
        &(-q,e^{\pm 2i\theta};-q)_\infty=\prod_{n=0}^\infty\left(1-(-q)^{n+1}\right)\left(1-e^{2i\theta}(-q)^{n}\right)\left(1-e^{-2i\theta}(-q)^n\right)\ ,\\
        &E(\theta)=a\cos^2\theta\ , \qquad a=\frac{4\mathcal{J}}{1+q}\ .\\
    \end{split}
\end{equation}
The triple product appearing in the density of states can be rewritten in the following form:
\begin{equation}
    \begin{split}
        &(-q,e^{\pm 2 i \theta(E)};-q)_\infty=-\left(e^{i\theta}-e^{-i\theta}\right)^2 \prod_{n=1}^\infty (1-(-q)^n)^3\left[1-\frac{(-q)^n}{(1-(-q)^n)^2}\left(e^{i\theta}-e^{-i\theta}\right)^2\right]\\
        &=-(-q,-q)^3_\infty\left(e^{i\theta}-e^{-i\theta}\right)^2\exp\left[-\sum_{m=1}^\infty \frac{\zeta_{-q}(2m)}{m}\left(e^{i\theta}-e^{-i\theta}\right)^{2m}\right]\\
        &=4(-q,-q)_\infty^3\sin^2\theta\exp\left[-\sum_{m=1}^\infty\frac{(-4)^m}{m}\zeta_{-q}(2m)(\sin^2\theta)^m\right]\\
        &=4(-q,-q)_\infty^3\frac{a-E}{a}\exp\left[-\sum_{m=1}^\infty \frac{(-4)^m}{m}\zeta_{-q}(2m)\left(\frac{a-E}{a}\right)^m\right]\ .\\
    \end{split}
\end{equation}
Using this along with the fact that $\sin(2\theta)=2\sin\theta\cos\theta=2a^{-1}\sqrt{E(a-E)}$ we have the following final expression for the $\mathcal{N}=1$ DSSYK model explicitly in terms of $E$:
\begin{equation}
\label{N=1DSSYKFullDOS}
    \rho_0(E)=\frac{2(-q,-q)^3_\infty}{\pi a}\sqrt{\frac{a-E}{E}}\exp\left[-\sum_{m=1}^\infty \frac{(-4)^m}{m}\zeta_{-q}(2m)\left(\frac{a-E}{a}\right)^m\right].
\end{equation}
We can clearly see the density of states is supported in the interval $E\in(0,a]$ with a hard edge at $E=0$ and a soft edge at $E=a$. In particular, it is interesting to note that as $q\to 0$ we obtain the Marchenko–Pastur density~\cite{Pastur:1967zca,Livan_2018}, this density is the $\mathcal{N}=1$ analogue of the statement that the ordinary DSSYK density in the $q\to 0$ regime is the Wigner semi-circle. Again, it should be noted that the density we got here should not be regarded as the density of some kind of double scaled matrix model. To get double scaled matrix models we should either expand around $E=0$ or $E=a$ spectral edge. 

\subsection{Notion of $\mathcal{N}=1$ q-deformed volumes from the hard edge}
Here we will describe the construction of a double scaled density from the hard edge of the $\mathcal{N}=1$ DSSYK model. To begin, we observe that if we simply analyze the density written in Eq. (\ref{N=1DSSYKFullDOS}) around $E=0$  the leading contribution coming from the exponential piece has no $E$ dependence, and we end up with a density going as $1/\sqrt{E}$ multiplied by some complicated q-dependent pre-factor which is not a useful for our purposes. However, consider the following manipulation of the density in Eq. (\ref{N=1DSSYKFullDOS}):
\begin{equation}
\label{N=1DSSYKFullDOSConvientRep}
\begin{split}
    &\rho_0(E)=\mathcal{C}\sqrt{\frac{a-E}{E}}\exp\left[-\sum_{m=1}^\infty \frac{(-4)^m}{m}\zeta_{-q}(2m)\left(\frac{a-E}{a}\right)^m\right]\\
    &=\mathcal{C}\sqrt{\frac{a-E}{E}}\exp\left[-\sum_{m=1}^\infty \frac{(-4)^m}{m}\sum_{n=1}^\infty \frac{(-q)^{nm}}{\left(1-(-q)^n\right)^{2m}}\left(\frac{a-E}{a}\right)^m\right]\\
    &=\mathcal{C}\sqrt{\frac{a-E}{E}}\exp\left[\sum_{n=1}^\infty \ln\left[1-\frac{(-q)^n}{\left(1-(-q)^n\right)^2}\frac{4(E-a)}{a}\right]\right]\\
    &=\mathcal{C}\sqrt{\frac{a-E}{E}}\prod_{n=1}^\infty\left[1-\frac{4(-q)^n}{a\left(1-(-q)^n\right)^2}(E-a)\right] \\
    &=\mathcal{C}\sqrt{\frac{a-E}{E}}\prod_{n=1}^\infty \left[\frac{(1+(-q)^n)^2}{(1-(-q)^n)^2}-\frac{(-q)^n}{(1-(-q)^n)^2}\frac{4E}{a}\right]\\
    &=\mathcal{C}\prod_{n=1}^\infty\left(\frac{(1+(-q)^n)^2}{(1-(-q)^n)^2}\right)\sqrt{\frac{a-E}{E}}\prod_{n=1}^\infty\left[1-\frac{(-q)^n}{(1+(-q)^n)^2}\frac{4E}{a}\right]\\
    &=\mathcal{C}'\sqrt{\frac{a-E}{E}}\exp\left[-\sum_{m=1}^\infty \frac{\xi_q(2m)}{m}\left(\frac{4E}{a}\right)^m\right]\ .\\
\end{split}
\end{equation}
In the first line we simply took the q-dependent pre-factor in front of the $\mathcal{N}=1$ DSSYK density and called it $\mathcal{C}$. In the second line we insert the definition of $\zeta_{-q}(2m)$. In the third line we swap the order of the $n$ and $m$ series and recognize the series over $n$ can be expressed in terms of $\ln$. Lines 4-6 are straightforward algebraic manipulations and to obtain the final line we essentially take the exponential of the log of the product and then expand the log and swap the order of the series. One of the series is absorbed into the definition of $\xi_q(2m)$ which is given by Eq. (\ref{xiDefintion}) which we rewrite below for the reader's convenience:
\begin{equation}
    \xi_q(2m)=\sum_{n=1}^\infty\frac{(-q)^{nm}}{(1+(-q)^n)^{2m}}\ .
\end{equation}
Through these manipulations we have rewritten the original density given in Eq. (\ref{N=1DSSYKFullDOS}) in a form which is ``naturally centered'' at the hard edge at $E=0$ now the argument in the exponential is not a simple factor. The price we paid for doing this is the appearance of a new q-deforming function $\xi_q$. After a truncation of this full density near $E=0$ we find the following structure for a candidate double scaled density:
\begin{equation}
    \rho_0(\epsilon)=\frac{C'}{\sqrt{\epsilon}}\exp\left[-\sum_{m=1}^\infty \frac{\xi_q(2m)}{m}\left(\gamma \epsilon\right)^m\right]\ ,
\end{equation}
where just like in the non-supersymmetric case, $C'$ and $\gamma$ will be fixed later using the scaled $q\to 1$ limit. In particular, to analyze this limit we need to understand the behavior of $\xi_q(2m)$ as $q\to 1$. This is done in Appendix \ref{xiqLimitAppendixProof}. The key result is the following scaling limit as $q\to 1$:
\begin{equation}
    \lim_{q\to 1}(1-q)^{2m}\xi_q(2m)=(-1)^m\left(1-\frac{1}{2^{2m}}\right)\zeta(2m)\ .
\end{equation}
Using this, we study the following scaled limit of the density:
\begin{equation}
    \begin{split}
        &\lim_{q\to 1}(1-q)\rho_0((1-q)^2\epsilon)=\frac{C'}{\sqrt{\epsilon}}\lim_{q\to 1}\exp\left[-\sum_{m=1}^\infty \frac{(1-q)^{2m}\xi_q(2m)}{m}(\gamma\epsilon)^m\right]\\
        &=\frac{C'}{\sqrt{\epsilon}}\exp\left[-\sum_{m=1}^\infty \frac{(-1)^m(1-2^{-2m})\zeta(2m)}{m}(\gamma\epsilon)^m\right]\\
        &=\frac{C'}{\sqrt{\epsilon}}\cosh\left(\frac{\pi\sqrt{\gamma\epsilon}}{2}\right)\ ,\\
    \end{split}
\end{equation}
which is, up to some conventions, the spectral density of $\mathcal{N}=1$ super JT gravity. We will adopt conventions where we set $\gamma=4$ and $C'=(2\pi\hbar)^{-1}$ which reproduces the density we mentioned in Eq. (\ref{OurDensityForN=1qDefVols})\footnote{In these conventions the disk density of states of $\mathcal{N}=1$ SJT gravity is given by $\rho_0(E)=\frac{\cosh(\pi\sqrt{E})}{2\pi\hbar\sqrt{E}}$ which is a different convention when compared to works like~\cite{Johnson:2021jtsugra,Johnson:2021explore,Johnson:2020mwi,Johnson:2022wsr} but we adopt this because it is a useful convention when we compare to Do and Norbury's work~\cite{DoNorbury:2025qMirzakhani}. Also see footnote \ref{footnoteN=1Volumeconventions} for a more detailed discussion for this convention.}. This particular choice of convention will provide a very useful and direct comparison to Do and Norbury's construction in~\cite{DoNorbury:2025qMirzakhani} which we will discuss in the next subsection. 

However, before doing this we will apply the ODE method discussed in Appendix \ref{N=1SUSYODEMethod}, to compute the resulting volumes obtained from this density. The first step is to determine the $t_k$ and $\mu$ associated with the density in Eq. (\ref{OurDensityForN=1qDefVols}). This is easily done by expanding the density around $E=0$ and reading off the $C_k$ which we then use in relation between $C_k$ and $t_k$ given in Eq. (\ref{N=1CanonicalForm}) to find explicit expressions for $t_k$, the first few are listed below:
\begin{equation}
    \begin{split}
        &\mu=t_0=1\ , \qquad t_1=-2\xi_q(2)\ , \qquad t_2=3\left(\xi_q(2)^2-\xi_q(4)\right)\ ,\\
        &t_3=-\frac{10}{3}\left(\xi_q(2)^3-3\xi_q(2)\xi_q(4)+2\xi_q(6)\right)\ .\\
    \end{split}
\end{equation}
Plugging these results into Eq. (\ref{N=1ODEVolumesGammaEq0}) gives the following expressions for our candidate q-deformed super Weil--Petersson volumes which we denote as $V^{qSWP}_{g,n}(b_1,...,b_n)$:
\begin{equation}
\label{OurqDefSWPVolumes}
    \begin{split}
        &V^{qSWP}_{1,n}(b_1,..,b_n)=(-1)^n\frac{(n-1)!}{8}\ ,\\
        &V^{qSWP}_{2,1}(b_1)=-\frac{3}{256}\left(b_1^2-24\xi_q(2)\right)\ ,\\
        &V^{qSWP}_{2,2}(b_1,b_2)=\frac{9}{256}\left(b_1^2+b_2^2-32\xi_q(2)\right)\ ,\\
        &V^{qSWP}_{3,1}(b_1)=-\frac{15}{8192}b^4+\frac{63}{256}\xi_q(2)b^2-\frac{9}{128}\left(59\xi_q(2)^2+25\xi_q(4)\right)\ .\\
    \end{split}
\end{equation}

What we can clearly see from the expressions provided is that  $V^{qSWP}_{g,n}(b_1,..b_n)$ are polynomials of degree $2g-2$, where we take the grading $\text{deg}(\xi_q(2m))=2m$ and $\text{deg}(b_i)=1$. Furthermore, one can explicitly check they satisfy the crucial scaled $q\to 1$ limit:
\begin{equation}
    \lim_{q\to 1}(1-q)^{2g-2}V^{qSWP}_{g,n}(b_{1,q},..,b_{n,q})=V_{g,n}^{SWP}(b_1,..,b_n)\ , \qquad b_{i,q}=\frac{b_i}{1-q}\ .
\end{equation}
Making them viable candidates to be dubbed as ``q-deformed'' $\mathcal{N}=1$ super Weil--Petersson volumes.

\subsection{Relation to Do and Norbury's $\mathcal{N}=1$ construction}
In this section we will address a natural question. Do the candidate q-deformed super Weil--Petersson volumes we obtained from the density given by Eq. (\ref{OurDensityForN=1qDefVols}) correspond to the same volumes that Do and Norbury defined in~\cite{DoNorbury:2025qMirzakhani} (denoted as $\hat{V}_{g,n}(b_1,..,b_n)$)? In Appendix \ref{DoNorburyqSWPVolComputations} we explicitly use Do and Norbury's recursion relations to compute the first few volumes which are given by:
\begin{equation}
\label{DoNorburyqDefSWPVols}
    \begin{split}
        &\hat{V}_{1,1}(b_1)=\frac{1}{8}\ ,\\
        &\hat{V}_{0,3}(b_1,b_2,b_3)=0\ ,\\
        &\hat{V}_{1,2}(b_1,b_2)=\frac{1}{8}\ ,\\
        &\hat{V}_{2,1}(b_1)=\frac{3}{256}\left(b_1^2+24\zeta_q^{odd}(2)\right)\ ,\\
        &\hat{V}_{0,4}(b_1,b_2,b_3,b_4)=0\ ,\\
        &\hat{V}_{1,3}(b_1,b_2,b_3)=\frac{1}{4}\ ,\\
        &\hat{V}_{2,2}(b_1,b_2)=\frac{9}{256}\left(b_1^2+b_2^2+32\zeta_q^{odd}(2)\right)\ ,\\
        &\hat{V}_{3,1}(b_1)=\frac{15}{8192}b_1^4+\frac{63}{256}\zeta^{odd}_q(2)b_1^2+\frac{9}{128}\left(59\zeta^{odd}_q(2)^2+25\zeta^{odd}_q(4)\right)\ .\\
    \end{split}
\end{equation}
A simple comparison of the expressions in Eq. (\ref{OurqDefSWPVolumes}) and Eq. (\ref{DoNorburyqDefSWPVols}) leads us to conclude that the two notions of volume are different. However, they are structurally very close to each other. In particular, we observe the following identity that maps our volumes to those of Do and Norbury:
\begin{eqnarray}
    (-1)^nV^{qSWP}_{g,n}(b_1,..,b_n)\vert_{\xi_q(2m)\to (-1)^m\zeta_q^{odd}(2m)}=\hat{V}_{g,n}(b_1,..,b_n)\ .
\end{eqnarray}
This relation's origin lies in the structure of $\xi_q(2m)$ itself. In particular, one can take the series defining $\xi_q$ and decompose the series into two component series. One series has an index running through odd numbers, denoted $\xi^{odd}_q(2m)$, and the other running through even numbers, denoted $\xi^{even}_q(2m)$, as written below:
\begin{equation}
\begin{split}
    &\xi_q(2m)=\xi_q^{odd}(2m)+\xi_q^{even}(2m)\ ,\\
    &\xi_q^{odd}(2m)=\sum_{n=1}^\infty \frac{(-q)^{(2n-1)m}}{\left(1+(-q)^{2n-1}\right)^{2m}}\ ,\\
    &\xi_q^{even}(2m)=\sum_{n=1}^\infty \frac{(-q)^{2nm}}{\left(1+(-q)^{2n}\right)^{2m}}\ .\\
\end{split}
\end{equation}
The key observation is to note:
\begin{equation}
    \xi_q^{odd}(2m)=(-1)^m\zeta_q^{odd}(2m)\ , 
\end{equation}
where $\zeta^{odd}_q(2m)$ was defined by Do and Norbury~\cite{DoNorbury:2025qMirzakhani} to be the following series:
\begin{equation}
    \zeta_q^{odd}(2m)=\sum_{n \text{ odd}}\frac{q^{nm}}{\left(1-q^n\right)^{2m}}=\sum_{n=1}^\infty \frac{q^{(2n-1)m}}{\left(1-q^{2n-1}\right)^{2m}}\ .
\end{equation}
Furthermore, using the results of Appendix \ref{xiqLimitAppendixProof} we know that:
\begin{equation}
    \lim_{q\to 1}(1-q)^{2m}\xi^{even}_q(2m)=0\  \qquad  \text{for}\qquad m\in\mathbb{N}\ .
\end{equation}
Therefore, the $\xi_q^{even}$ will completely vanish and play no role in the scaling limits where we reproduce the usual super Weil--Petersson volumes. We will refer to $\xi^{even}_q(2m)$ as the ``irrelevant'' part of $\xi_q(2m)$. Throwing away the irrelevant part allows us to express everything in terms of Do and Norbury's $\zeta_q^{odd}(2m)$. In particular, we can see that the double scaled spectral density that reproduces their q-deformed $\mathcal{N}=1$ super Weil--Petersson volumes (up to a $(-1)^n$ convention that appears in the ODE method framework) is exactly given by taking the density in Eq. (\ref{OurDensityForN=1qDefVols}) and replacing $\xi_q(2m)\to (-1)^m\zeta^{odd}_q(2m)$ which yields:
\begin{equation}
    \rho^{DN}_0(E)=\frac{1}{2\pi\hbar\sqrt{E}}\exp\left[-\sum_{m=1}^\infty \frac{(-4)^m\zeta_q^{odd}(2m)}{m}E^m\right]\ .
\end{equation}

Based on the analysis done here we conclude that it is indeed possible to arrive at the at the volumes defined by Do and Norbury using the local structure of the hard edge of $\mathcal{N}=1$ DSSYK model spectral density. Although the exact procedure is slightly more subtle than it was in non-supersymmetric case the overall idea that Do and Norbury's $\mathcal{N}=1$ volumes derive from the edge structure of $\mathcal{N}=1$ DSSYK model remains intact.

\subsection{Discrete analogue of super WP volumes from $\mathcal{N}=1$ DSSYK?}
\label{N=1DiscreteWPVolSec}
Now that we have a clear understanding of how to think of Do and Norbury's volumes in terms of the $\mathcal{N}=1$ DSSYK model we can also ask if these supersymmetric q-deformed volumes appear as highest degree terms of some $\mathcal{N}=1$ analogue of Okuyama's discrete Weil--Petersson volumes~\cite{Okuyama:2023discrete}. 

To explore this question we will return to the full $\mathcal{N}=1$ DSSYK density of states in Eq. (\ref{N=1DSSYKFullDOS}) and its alternate representation in Eq. (\ref{N=1DSSYKFullDOSConvientRep}). Using these densities of state we can attempt to write down an expression for a spectral curve similar to how Okuyama defined the spectral curve in Eq. (\ref{eq:Do-Norbury-full-curve}) from the spectral density given in Eq. (\ref{DSSYKFullDOS}). 

Essentially we need to find a uniformizing map, $E=E(z)$, such that the the density of states of $z$ has no branch points in $z$ variables. This can be achieved by the following map:
\begin{equation}
    E(z)=\frac{a}{2}+\frac{a}{4}(z+z^{-1})\ .
\end{equation}
The density of states in Eq. (\ref{N=1DSSYKFullDOS}) up to some factor $C$ is takes on the structure:
\begin{equation}
    \rho_0(z)=C \left(\frac{z-1}{z+1}\right)\exp\left[-\sum_{m=1}^\infty \frac{\zeta_{-q}(2m)}{m}\frac{(z-1)^{2m}}{z^m}\right]\ .
\end{equation}
Based on this we can consider a spectral curve of the form (where we take $E(z)\propto x(z)$):
\begin{equation}
    \begin{split}
    \label{N=1SpecCurve}
        &x(z)=\frac{a}{2}+\frac{a}{4}(z+z^{-1})\ ,\\
        &y(z)=C\left(\frac{z-1}{z+1}\right)\exp\left[-\sum_{m=1}^\infty \frac{\zeta_{-q}(2m)}{m}\frac{(z-1)^{2m}}{z^m}\right]\ .\\
    \end{split}
\end{equation}
There a number of important observations to make. The first is that there are two ramification points at $z=\pm 1$. We can see that $z=-1$ corresponds to $x(z)=E=0$ which is the location of the hard edge of the density of states and ramification point at $z=+1$ corresponds to the soft edge at $E=a$. Furthermore, we can also see that the argument in the exponential only vanishes at the soft edge and not the hard edge at $z=-1$. This will have important consequences when apply the machinery of topological recursion to generate $\omega_{g,n}(z_1,..,z_n)$ since these are constructed by taking residues around the ramification points at $z=\pm 1$. In particular, for the curve given in Eq. (\ref{N=1SpecCurve}) we know the involution map is given by $\sigma(z)=z^{-1}$. One can also easily check that $y(1/z)=-y(z)$. It follows that (assuming standard Bergman kernel i.e. $\omega_{0,2}(z_1,z_2)=\frac{dz_1dz_2}{(z_1-z_2)^2}$) the recursion kernel is given by the following expression:
\begin{equation}
    \begin{split}
        &K(z_1,z)=\frac{\frac{1}{z_1-z}-\frac{1}{z_1-1/z}}{4y(z)\frac{dx}{dz}}\frac{dz_1}{dz}=\frac{z^2}{a(z_1-z)(zz_1-1)y(z)}\ .\\
    \end{split}
\end{equation}
This allows us to compute $\omega_{g,n}$. For example, the computation of $\omega_{1,1}(z_1)$ will proceed as follows:
\begin{equation}
    \begin{split}
    &\omega_{1,1}(z_1)=\mathop{\rm Res}_{z\to -1}
    \left[-K(z_1,z)\right]\omega_{0,2}(z,1/z)+\mathop{\rm Res}_{z\to +1}
    \left[-K(z_1,z)\right]\omega_{0,2}(z,1/z)\\
    &=\left[\frac{\exp\left[\sum_{m=1}^\infty \frac{(-4)^m\zeta_{-q}(2m)}{m}\right]}{8aC(1+z_1)^2}+\frac{4(z_1-1)^2\zeta_{-q}(2)+z_1(6-z_1)-1}{8aC(z_1-1)^4}\right]dz_1\ .\\
    \end{split}
\end{equation}
At this level in the calculation we immediately see a problem. Namely that, if we assign a grading $\text{deg}(\zeta_{-q}(2m))=2m$ we can see that the $\exp$ term has an ill defined degree. The origin of the $\exp$ term can be traced back to taking the residue at $z=-1$ (the hard edge) the argument in the $\exp$ does not vanish at $z=-1$ (this is in contrast to the soft edge at $z=1$ in which the argument of the exponential does vanish and we see the residue gives an expression which would have a well defined degree). One may attempt to remedy this issue by suggesting that one should really be finding the spectral curve associated to the alternate form of the density in Eq. (\ref{N=1DSSYKFullDOSConvientRep}) since it is centred around the hard edge. In this case, the expression for $x(z)$ remains unchanged but $y(z)$ is slightly different. However, the problem still persists the problematic terms in this representation come from the soft edge at $z=1$ where the argument of the exponential will not vanish. In either case, there appears to be a problem in finding a $\mathcal{N}=1$ analogue of discrete Weil--Petersson volumes from the $\mathcal{N}=1$ DSSYK model which naturally contains the q-deformed $\mathcal{N}=1$ volumes as a highest degree contribution.     

\section{Discussion and conclusion}
\label{DiscussionConclusionSection}
In this work, we have explored the idea of interpreting the recent constructions of q-deformed Weil--Petersson volumes discussed in the works~\cite{Okuyama:2023discrete,DoNorbury:2025dssyk,DoNorbury:2025qMirzakhani} using the DSSYK and its $\mathcal{N}=1$ analogue as starting points. In both cases we have demonstrated a procedure to construct q-deformed volumes by exploiting the structure of the spectral density edges of DSSYK models. 

In Section \ref{qWPVolFromEdgeDSSYKSection}, we showed that the q-deformed volumes of Do and Norbury~\cite{DoNorbury:2025dssyk,DoNorbury:2025qMirzakhani} which also appear as highest degree terms in Okuyama's discrete volumes~\cite{Okuyama:2023discrete} can be constructed from purely local data of the soft spectral edge of the DSSYK model. We also identified a specific double scaled matrix model spectral density given by Eq. (\ref{qDefVolDOS}) which exactly reproduces the q-deformed Weil--Petersson volumes (as shown in sample calculations of Appendix \ref{sec:top-first-volumes}) using the machinery of the ODE method discussed in Appendix \ref{sec:ODE-method}. In Section \ref{N=1DSSYKWPVolSection}, we explored the possibility of defining $\mathcal{N}=1$ q-deformed super Weil--Petersson volumes by zooming in on the edges of the $\mathcal{N}=1$ DSSYK model. In particular, we observed that the $\mathcal{N}=1$ DSSYK model had one hard edge at $E=0$ and one soft edge at $E=a$. By focusing on the structure of the hard edge we identified the spectral density given by Eq. (\ref{OurDensityForN=1qDefVols}) which involved a q-series, denoted $\xi_q$. Using the ODE method we obtained candidate expressions for $\mathcal{N}=1$ q-deformed super Weil--Petersson volumes, given in Eq. (\ref{OurqDefSWPVolumes}), and demonstrated that they are related Do and Norbury's $\mathcal{N}=1$ q-deformed volumes (given in Eq. (\ref{DoNorburyqDefSWPVols})) through the replacement $\xi_q(2m)\to(-1)^m\zeta_q^{odd}(2m)$ and an overall sign $(-1)^n$. The origin of this replacement lies in the structure of $\xi_q(2m)$ and by splitting it into odd and even index series we find the odd part reproduces the combination $(-1)^m\zeta_q^{odd}(2m)$ and the even part represents a contribution which goes to zero in the scaled $q\to 1$ limit. If we discard the even sector, we obtain the density in Eq. (\ref{DoNorburyVolumesDOS}), thereby demonstrating the $\mathcal{N}=1$ DSSYK model hard edge origin of Do and Norbury's $\mathcal{N}=1$ volumes.

Together these results suggest a more broader picture that q-deformed volumes of various types can be constructed quite readily by using the edge structure of some corresponding DSSYK model. In particular, if we go back to the $\mathcal{N}=1$ DSSYK model we saw that it had two types of edges one was the hard edge at $E=0$ which essentially gave us Do and Norbury's $\mathcal{N}=1$ volumes in terms of $\zeta_q^{odd}(2m)$. What should we expect from the soft edge of this model? Intuitively, we should expect some kind of q-deformation of the Weil--Petersson volumes to emerge. As it turns out from the analysis in Appendix \ref{N=1SoftEdgeVolumes} this intuition is in fact true. We obtain a density with a structure that is similar to what we had analyzing the soft edge of the non-supersymmetric DSSYK model. However, instead of involving $\zeta_q(2m)$ it now involves $\zeta_{q^2}(2m)=\zeta_q^{even}(2m)$ which is an interesting result. Namely the hard edge of $\mathcal{N}=1$ involves $\zeta_q^{odd}(2m)$ and the soft edge of the same model involves the complimentary q-series $\zeta_q^{even}(2m)$. 

An open question which we posed in Section \ref{N=1DiscreteWPVolSec} is whether the $\mathcal{N}=1$ q-deformed volumes have some type of interpretation in terms of the highest degree terms of some discrete analogue of $\mathcal{N}=1$ super Weil--Petersson volumes. To answer this question we employed an approach similar to Okuyama in~\cite{Okuyama:2023discrete}. After finding the associated spectral curve to $\mathcal{N}=1$ DSSYK we computed $\omega_{1,1}(z)$ and demonstrated that it will contain terms with an ill defined degree. We do not know whether this is a fundamental property of $\mathcal{N}=1$ theories or an artifact of our naive generalization of the bosonic case\footnote{In a companion work by a subset of the authors~\cite{Ahmed:2026rlc}, we have studied $\mathcal{N}=2$ DSSYK and the corresponding $q$-deformed spectral curves. Interestingly, the full discrete curve gives well-behaved quasi-polynomials, and the local edge expansions give the bosonic and $\mathcal{N}=2$ top-degree polynomial volumes, respectively. In particular, the lower edge, which retains the pole associated with the BPS states, gives $q$-deformed $\mathcal{N}=2$ super Weil--Petersson volumes that reduce to those of Turiaci and Witten~\cite{Turiaci:2023jfa} in the continuum limit.}. In the latter case, Okuyama's approach may need to be modified in the presence of a hard edge to properly arrive at some notion of $\mathcal{N}=1$ analogues of discrete super Weil--Petersson volumes which also nicely embeds the $\mathcal{N}=1$ q-deformed volumes of Do and Norbury~\cite{DoNorbury:2025qMirzakhani}. Perhaps the observation we made for $\mathcal{N}=1$ DSSYK, that the soft edge involves $\zeta_q^{even}(2m)$ and hard edge involves $\zeta_q^{odd}(2m)$ will be central in such constructions?  

Another interesting question we can ask pertains to how the hard edge of $\mathcal{N}=1$ DSSYK model naturally gave rise to the $\xi_q$ rather than $\zeta_q^{odd}$ defined in Do and Norbury's construction. This leads to the question if there is some physical interpretation to the irrelevant terms in $\xi_q$. Are they just some kind of model dependent artifacts from the DSSYK model or do they encode some kind of important physics at finite $0<q<1$ regime?  

Aside from these immediate open problems/questions we have successfully written the spectral density of a double scaled matrix model which reproduce the volumes rigorously defined by Do and Norbury in~\cite{DoNorbury:2025qMirzakhani} and also provided a way to construct them from the edge structure of various DSSYK models. This opens up a larger landscape of exploration using tools that have been recently developed~\cite{Johnson:2026jbq,Johnson:2026twg,Johnson:2026jls,Johnson:2026jgs,Johnson:2026wlo} which may provide insight on understanding non-perturbative aspects of q-deformed volumes and beyond~\cite{Johnson:2026jbq,Alfinito:2026pst,Ahmed:2026idm}.

\section*{Acknoweldgements}
J.N. is supported by the Beijing Institute of Mathematical Sciences and Applications. This project was supported in part by grant NSF PHY-2309135 to the Kavli Institute for Theoretical Physics (KITP). J.N. would like to thank the KITP for hospitality during the Graduate Fellowship program during which this research was conducted. Krishan Saraswat is supported by the University of California Santa Barbara and the US Department of Energy under the grant DE-SC 0011687. W.A. is supported by the U.S. Department of Energy under grant DE-SC 0011702. The authors would like to thank Clifford Johnson and Jiuci Xu for helpful discussions and feedback.  

\appendix

\section{Overview of important formalisms}
\label{sec:overview}
\subsection{Topological recursion}
\label{sec:topological-recursion}

This section reviews the version of topological recursion that will be discussed throughout the paper. This is the current standard technique used to calculate the invariants. The construction is the Chekhov--Eynard--Orantin recursion~\cite{Chekhov:2006vd,Eynard:2007kz}, specialized below to the genus-zero spectral curves that will be needed later. The starting point is a spectral curve:
\begin{equation}
    \cS=(\Sigma,x,y,B)\ ,
\end{equation}
consisting of a Riemann surface $\Sigma$, two meromorphic functions $x$ and $y$, and a symmetric bidifferential $B$. In a global coordinate $z$ on $\Sigma\simeq\mathbb{CP}^{1}$, the basic recursion data are:
\begin{equation}
\label{eq:TR-initial-data}
    x=x(z)\ ,\qquad
    \omega_{0,1}(z)=y(z)dx(z)\ ,\qquad
    \omega_{0,2}(z_1,z_2)=B(z_1,z_2)=\frac{dz_1dz_2}{(z_1-z_2)^2}\ .
\end{equation}

The ramification points are the zeros of $dx$. We will assume that they are simple, and denote their set by:
\begin{equation}
\label{eq:ramification-set}
    \Ram=\{a\in\Sigma\,|\,dx(a)=0\}\ .
\end{equation}
For each $a\in\Ram$, there is a local involution map satisfying the following properties:
\begin{equation}
\label{eq:local-involution}
    x(z)=x(\sigma_a(z))\ ,\qquad \sigma_a(z)\neq z\ ,\qquad \sigma_a(a)=a\ .
\end{equation}
Using this data, we can define the recursion kernel in a neighborhood of $a$ by:
\begin{equation}
\label{eq:TR-kernel}
    K_a(z_1,z)=
    \frac{\displaystyle\int_{\sigma_a(z)}^z B(z_1,\zeta)}
    {2\left[\omega_{0,1}(z)-\omega_{0,1}(\sigma_a(z))\right]}\ .
\end{equation}
In our case, the numerator can be evaluated explicitly:
\begin{equation}
\label{eq:TR-kernel-numerator}
    \int_{\sigma_a(z)}^z B(z_1,\zeta)
    =\left[\frac{1}{z_1-z}-\frac{1}{z_1-\sigma_a(z)}\right]dz_1\ .
\end{equation}
Equation~(\ref{eq:TR-kernel}) becomes:
\begin{equation}
\label{eq:TR-kernel-explicit}
    K_a(z_1,z)=
    \frac{\left[\frac{1}{z_1-z}-\frac{1}{z_1-\sigma_a(z)}\right]dz_1}
    {2\left[y(z)-y(\sigma_a(z))\right]dx(z)}\ .
\end{equation}
For every stable pair $(g,n)$, {
\it i.e.,} $2g-2+n>0$, the correlators $\omega_{g,n}$ are determined recursively by:
\begin{eqnarray}
\label{eq:topological-recursion}
\omega_{g,n}(z_1,Z)
&=&\sum_{a\in\Ram}\mathop{\rm Res}_{z\to a}\left[-K_a(z_1,z)\right]
\left\{\omega_{g-1,n+1}(z,\sigma_a(z),Z)\right.\nonumber\\
&&\left.\hspace{1.1cm}
+\sum_{\substack{g_1+g_2=g\\I\sqcup J=Z}}^{\prime}
\omega_{g_1,|I|+1}(z,I)
\omega_{g_2,|J|+1}(\sigma_a(z),J)
\right\}\ ,
\end{eqnarray}
where $Z=(z_2,\ldots,z_n)$. The prime means that terms containing an unstable factor $\omega_{0,1}$ are omitted. Our convention places an explicit minus sign in front of the kernel. This is convenient for the volume conventions used below, and avoids carrying an additional factor $(-1)^\chi$, where $\chi=2-2g-n$. It is useful to display the first few cases to see how the formula works. At genus one with one boundary:
\begin{equation}
\label{eq:omega11-TR}
    \omega_{1,1}(z_1)=
    \sum_{a\in\Ram}\mathop{\rm Res}_{z\to a}
    \left[-K_a(z_1,z)\right]\omega_{0,2}(z,\sigma_a(z))\ .
\end{equation}
The three-boundary sphere amplitude is:
\begin{eqnarray}
\label{eq:omega03-TR}
\omega_{0,3}(z_1,z_2,z_3)
&=&\sum_{a\in\Ram}\mathop{\rm Res}_{z\to a}
\left[-K_a(z_1,z)\right]
\left\{\omega_{0,2}(z,z_2)\omega_{0,2}(\sigma_a(z),z_3)\right.\nonumber\\
&&\left.\hspace{4.5cm}
+\omega_{0,2}(z,z_3)\omega_{0,2}(\sigma_a(z),z_2)\right\}\ .
\end{eqnarray}
We also have:
\begin{eqnarray}
\label{eq:omega12-TR}
\omega_{1,2}(z_1,z_2)
&=&\sum_{a\in\Ram}\mathop{\rm Res}_{z\to a}
\left[-K_a(z_1,z)\right]
\left\{\omega_{0,3}(z,\sigma_a(z),z_2)\right.\nonumber\\
&&\left.\hspace{2.0cm}
+\omega_{1,1}(z)\omega_{0,2}(\sigma_a(z),z_2)
+\omega_{1,1}(\sigma_a(z))\omega_{0,2}(z,z_2)\right\}\ ,
\end{eqnarray}
and
\begin{equation}
\label{eq:omega21-TR}
    \omega_{2,1}(z_1)=
    \sum_{a\in\Ram}\mathop{\rm Res}_{z\to a}
    \left[-K_a(z_1,z)\right]
    \left\{\omega_{1,2}(z,\sigma_a(z))
    +\omega_{1,1}(z)\omega_{1,1}(\sigma_a(z))\right\}\ .
\end{equation}
The four-boundary sphere amplitude can be written as:
\begin{eqnarray}
\label{eq:omega04-TR}
\omega_{0,4}(z_1,z_2,z_3,z_4)
&=&\sum_{a\in\Ram}\mathop{\rm Res}_{z\to a}
\left[-K_a(z_1,z)\right]
\sum_{i=2}^{4}\bigg\{
\omega_{0,2}(z,z_i)\omega_{0,3}(\sigma_a(z),Z_i)
\nonumber\\
&&\hspace{4.8cm}
+\omega_{0,3}(z,Z_i)\omega_{0,2}(\sigma_a(z),z_i)
\bigg\}\ ,
\end{eqnarray}
where $Z_i=\{z_2,z_3,z_4\}\setminus\{z_i\}$.

The important structural fact is that every $\omega_{g,n}$ is obtained from residues localized at the ramification points. Consequently, topological recursion is sensitive not only to the local expansion of $y$ but also to the full ramification structure and to the bidifferential $B$. 

In certain contexts the inverse Laplace transform of the form generated by topological recursion have the interpretation of  ``volumes'' $V_{g,n}(b_1,..,b_n)$. More precisely we have:
\begin{equation}
\label{eq:inverse-laplace-volumes}
    V_{g,n}(b_1,\ldots,b_n)
    =\frac{1}{b_1\cdots b_n}
    \cL^{-1}_{(z_1,..,z_n)}
    \left(\frac{\omega_{g,n}(z_1,\ldots,z_n)}{\prod_{i=1}^n dz_i}\right)[b_1,..,b_n]\ .
\end{equation}
\subsection{The ODE method to produce volumes}
\label{sec:ODE-method}
As we mentioned in the introduction of this work the ODE method to produce volumes revolves around an auxiliary Schrodinger problem in Eq. (\ref{SchrodingerEq}) where $u(x)$ is a solution to the string equation given in Eq. (\ref{FullStringEquation}). In this appendix we will review essential formulas needed to construct the volumes discussed in this paper. In particular, a key relation that relates the zeroth order (in the perturbative $\hbar$ expansion) solution to the string equation to the disk level density is:  
\begin{equation}
\label{eq:disk-density}
    \rho_0(E)=\frac{1}{2\pi\hbar}
    \int_{-\infty}^{\mu}
    \frac{\Theta(E-u_0(x))}{\sqrt{E-u_0(x)}}dx\ .
\end{equation}
Under the assumption that $u_0(x)$ is a monotonically deceasing function and that $u_0(\mu)=E_0>0$. In such a case we can eliminate the Heaviside step function and give the lower bound of the integral as $x_*<\mu$ where $E-u_0(x_*)=0$ (where $E>E_0$) and write:
\begin{equation}
    \rho_0(E)=\frac{1}{2\pi\hbar}\int_{x_*}^\mu \frac{1}{\sqrt{E-u_0}}dx\ .
\end{equation}
The tree level string equation is the lowest order set of terms in the perturbative $\hbar$ expansion of Eq. (\ref{FullStringEquation}) which is explicitly given as:
\begin{equation}
    u_0(x)\left[\sum_{k=1}^\infty t_k u_0(x)^k+x\right]^2=0\ .
\end{equation}
We can see there are two distinct types of solutions one in which $u_0(x)=0$ and the other where the expression in the square bracket vanishes. Which solutions we consider will depend if we are considering non-supersymmetric or supersymmetric models.
\subsubsection{Non-supersymmetric context}
For the purposes of obtaining Weil--Petersson volumes and their q-deformations we will be interested in solutions over the domain $x<0$ where $u_0(x)\neq 0$. In this case the tree level equation becomes:
\begin{equation}
\label{eq:tree-string-equation}
    G(u_0(x))+x=0\ ,\qquad
    G(u_0)=\sum_{k=1}^{\infty}t_k u_0^k\ .
\end{equation}
Using this tree level string equation we can introduce a change of integration variables from $x$ to $u_0$ which enables us to write $\rho_0$ in terms of the following integral:
\begin{equation}
    \begin{split}
        &\rho_0(E)=\frac{1}{2\pi\hbar}\int_{E}^{E_0}\frac{-\sum_{k=1}^\infty kt_k u_0^{k-1}}{\sqrt{E-u_0}}du_0=\frac{1}{2\pi\hbar}\sum_{k=1}^\infty kt_k\int_{E_0}^E\frac{u_0^{k-1}}{\sqrt{E-u_0}}du_0\ .\\
    \end{split}
\end{equation}

The definite integral above can be done exactly to give the following result:
\begin{equation}
    \begin{split}
        &\rho_0(E)=\frac{1}{\pi\hbar}\sum_{k=1}^\infty \sum_{p=0}^{k-1} k t_k E_0^{k-1/2} q_{p,k}\lambda^{p+1/2}\ ,\\
        &\lambda=\frac{E}{E_0}-1\ ,\\
        &q_{p,k}=\binom{k-1}{p}\frac{2^p p!}{(2p+1)!!}\ .\\
    \end{split}
\end{equation}
One can then swap the order between the sum and series to obtain the following double series expression\footnote{One can check that the swapping can be done via the formula $\sum_{k=n_*}^\infty \sum_{p=n_*}^ka_{p,k}=\sum_{p=n_*}^\infty \sum_{k=p}^\infty a_{p,k}$, where we set $n_*=1$.}:
\begin{equation}
    \rho_0(E)=\frac{1}{\pi\hbar}\sum_{p=1}^\infty \left(\sum_{k=p}^\infty kt_kE_0^{k-p}q_{p-1,k}\right)(E-E_0)^{p-1/2}\ .
\end{equation}
Now we have a representation of the form (we will refer to this as the ``canonical'' representation):
\begin{equation}
\begin{split}
    \label{eq:canonical-density}
    &\rho_0(E)=\frac{1}{\pi\hbar} \sum_{p=1}^\infty C_p (E-E_0)^{p-1/2}\ ,\\
    &C_p=\sum_{k=p}^\infty kt_kE_0^{k-p}q_{p-1,k}=\frac{2^{p-1}}{(2p-1)!!}\frac{d^p}{dE_0^p}\sum_{k=p}^\infty t_k E_0^k\ .\\
\end{split}
\end{equation}
This gives a relation between $t_k$ and $C_p$. In the ODE approach we are handed a $\rho_0(E)$ (i.e. $C_p$) and our task is to determine the associated $t_k$. This is achieved by inverting the relation above. One can check that the following relation between $t_k$ and $C_k$ does the job:
\begin{equation}
\label{eq:tk-inversion}
    t_k=\frac{1}{k!}\sum_{\ell=k}^\infty \frac{(-E_0)^{\ell-k}}{(\ell-k)!}\frac{(2\ell-1)!!}{2^{\ell-1}}C_\ell\ .
\end{equation} 
For the purposes of this work it will suffice to set $E_0=0$. In which case we have a much more simple relation between $t_k$ and $C_k$:
\begin{equation}
\label{simpletkCkrelation}
    t_k=2\frac{\binom{2k}{k}}{4^k}C_k\ .
\end{equation}

This will give the associated choice of $t_k$ for a given density of states that can be expressed in the canonical form of Eq. (\ref{eq:canonical-density}). Once we have fixed the $t_k$ through this procedure we have uniquely identified the string equation for that model which enables us to understand the behavior of $u_0$ at the Fermi surface. In particular, we can then differentiate the string equation iteratively and compute expressions for the derivative of $u_0(x)$ at any point $x$. In particular, the key point for us will be $x=\mu$ where $u_0(\mu)=E_0$ which is all we will need to construct $W_{g,n}$ using the results described in~\cite{Johnson:2026twg}. For example, the expression for $W_{1,1}(z)$ in terms of the derivatives of $u_0$ is~\cite{Johnson:2026twg}:
\begin{equation}
\label{W11Formula}
    W_{1,1}(z)=\frac{1}{24}\frac{u''_0(\mu)}{u_0'(\mu)}\frac{1}{z^2}-\frac{u_0'(\mu)}{16}\frac{1}{z^4}\ .\\
\end{equation}
Then using relation:
\begin{equation}
\label{eq:inverse-laplace-volumes}
    V_{g,n}(b_1,\ldots,b_n)
    =\frac{1}{b_1\cdots b_n}
    \cL^{-1}_{(z_1,..,z_n)}
    \left(W_{g,n}(z_1,\ldots,z_n)\right)[b_1,..,b_n]\ ,
\end{equation}
will yield the volume $V_{1,1}(b)$.

In the cases where $E_0=0$ a key observation about the tree level string equation in Eq. (\ref{eq:tree-string-equation}) is how setting $E_0=0$ implies that the Fermi-surface is located at $\mu=0$\footnote{This can be seen by noting that $u_0(x=\mu)=E_0=0$ so then the tree level string equation at the Fermi surface reads $\sum_{k=1}^\infty t_k\cdot 0^k+\mu=0\Rightarrow \mu=0$.}. Using this fact we can then write the solution to the string equation in a neighborhood of $x=\mu=0$ as:
\begin{equation}
    u_0(x)=\sum_{p=1}^\infty \frac{u_0^{(p)}(0)}{p!}x^p\ . 
\end{equation}
We can plug this Anzatz back into Eq. (\ref{eq:tree-string-equation}) and expand around $x=0$ and solve the equations order by order. The end result will be expressions for the derivatives of $u_0(0)$ in terms of the $t_k$ we computed from the disk density of states. For example, for the first 5 derivatives we will find:
\begin{equation}
\label{5u0primetExpression}
\begin{split}
    &u_0'(0)=-\frac{1}{t_1}\ ,\\
    &u_0''(0)=-\frac{2t_2}{t_1^3}\ ,\\
    &u_0^{(3)}(0)=\frac{6\left(-2t_2^2+t_1t_3\right)}{t_1^5}\ ,\\
    &u_0^{(4)}(0)=-\frac{24\left(5t_2^3-5t_1t_2t_3+t_t^2t_4\right)}{t_1^7}\ ,\\
    &u_0^{(5)}(0)=\frac{120\left[-14 t_2^4 + 21 t_1 t_2^2 t_3 - 6 t_1^2 t_2 t_4 + t_1^2 (-3 t_3^2 + t_1 t_5)\right]}{t_1^9}\ .\\
\end{split}
\end{equation}
We can directly apply these formulas in Appendix \ref{sec:top-first-volumes} to demonstrate that ODE method reproduces the q-deformed Weil--Petersson volumes when $\rho_0(E)$ is given by Eq. (\ref{qDefVolDOS}).  \\

\subsubsection{$\mathcal{N}=1$ supersymmetric context}
\label{N=1SUSYODEMethod}
Typically in the case of $\mathcal{N}=1$ supersymmetry, the solutions of interest is piecewise defined. When $x<0$ it is assumed that $u_0(x)\neq 0$ and when $x>0$ we assume that $u_0(x)>0$ we also take the location of the Fermi-surface $x=\mu>0$. Using this, we can split the integral describing the density as follows:
\begin{equation}
\begin{split}
     &\rho_0(E)=\frac{1}{2\pi\hbar}\left(\int_{-\infty}^0\frac{\Theta(E-u_0(x))}{\sqrt{E-u_0(x)}}dx+\int_0^\mu\frac{\Theta(E)}{\sqrt{E}}dx\right)\\
    &=\frac{1}{2\pi\hbar}\left(\frac{\Theta(E)\mu}{\sqrt{E}}+\int_{-\infty}^0\frac{\Theta(E-u_0(x))}{\sqrt{E-u_0(x)}}dx\right)\ .\\
\end{split}
\end{equation}
The second integral can be treated in analogy to what we did in the non-supersymmetric discussion (we will assume $u_0(x)$ decreases monotonically and vanishes at $x=0$ so it continuously connects $x>0$ solution). Then we can write:
\begin{equation}
    \rho_0(E)=\frac{1}{2\pi\hbar}\left(\frac{\mu}{\sqrt{E}}+\frac{1}{\sqrt{E}}\sum_{k=1}^\infty t_k\frac{4^k}{\binom{2k}{k}}E^{k}\right),\qquad E>0\ .
\end{equation}
Therefore the density that we get can be written in the following ``canonical'' form:
\begin{equation}
\begin{split}
\label{N=1CanonicalForm}
    &\rho_0(E)=\frac{\sum_{k=0}^\infty C_k E^k}{\pi\hbar\sqrt{E}}\ ,\\
    &\mu=2C_0\ ,\qquad t_k=2\frac{\binom{2k}{k}}{4^k}C_k\ .\\
\end{split}
\end{equation}
In analogy to how knowledge of $t_k$ allows one to write expressions for $W_{g,n}$ in terms of $t_k$ a similar story also holds in the supersymmetric context. For example, we have\footnote{In the work~\cite{Johnson:2026twg} the formulas that were written assumed that $\mu=1$. Here we are writing those formulas without that assumption and restoring explicit $\mu$ dependence.}:
\begin{equation}
    \begin{split}
    &W_{1,n}(z_1,..,z_n)=\frac{(-1)^{n-1}(n-1)!\left(\Gamma^2-\frac{1}{4}\right)}{2\mu^n}\prod_{i=1}^n\frac{1}{z_i^2}\ ,\\
    &W_{2,1}(z)=-\left(\Gamma^2-\frac{1}{4}\right)\left(\Gamma^2-\frac{9}{4}\right)\left[\frac{1}{8\mu^3 z^4}+\frac{t_1}{4\mu^4z^2}\right]\ ,\\
    &W_{2,2}(z_1,z_2)=\left(\Gamma^2-\frac{1}{4}\right)\left(\Gamma^2-\frac{9}{4}\right)\left[\frac{t_1}{\mu^5z_1^2z_2^2}+\frac{3(z_1^2+z_2^2)}{8\mu^4z_1^4z_2^4}\right]\ ,\\
    &W_{3,1}(z)=\frac{1}{240}\left(\Gamma^2-\frac{1}{4}\right)\left(\Gamma^2-\frac{9}{4}\right)\Biggl[\frac{20\left[6\left(\Gamma^2-\frac{21}{4}\right)t_1^2-2\mu\left(\Gamma^2-\frac{25}{4}\right)t_2\right]}{\mu^7 z^2}\\
    &\hspace{3.4cm}+\frac{60\left(\Gamma^2-\frac{21}{4}\right)t_1}{\mu^6z^4}+\frac{15\left(\Gamma^2-\frac{25}{4}\right)}{\mu^5z^6}\Biggr]\ .\\
    \end{split}
\end{equation}
The volumes are then simply computed from the $W$'s using the relation in Eq. (\ref{eq:inverse-laplace-volumes}). In the case where we set $\Gamma=0$ (which is the case we will primarily be interested in which corresponds to a pure NS boundary sector) we get: 
\begin{equation}
\label{N=1ODEVolumesGammaEq0}
    \begin{split}
        &V_{1,n}(b_1,..,b_n)=\frac{(-1)^n(n-1)!}{8\mu^n}\ ,\\
        &V_{2,1}(b)=-\frac{3\left(b^2+\frac{12t_1}{\mu}\right)}{256\mu^3}\ ,\\
        &V_{2,2}(b_1,b_2)=\frac{9\left(b_1^2+b_2^2+\frac{16t_1}{\mu}\right)}{256\mu^4}\ ,\\
        &V_{3,1}(b)=-\frac{15b^4+\frac{1008t_1}{\mu}b^2+192\left(\frac{63t_1^2}{\mu^2}-\frac{25t_2}{\mu}\right)}{8192\mu^5}\ .\\
    \end{split}
\end{equation}
The formulas above can directly be used to obtain the expressions for volumes obtained from the density of state given by Eq. (\ref{OurDensityForN=1qDefVols}) and Eq. (\ref{DoNorburyVolumesDOS}) once the appropriate $t_k$ are computed we can use them to obtain the results of Eq. (\ref{OurqDefSWPVolumes}).

\subsection{Translating topological recursion approach to ODE approach}
\label{sec:matching-ODE-TR}
The class of curves we wish to consider in this work are of the form:
\begin{equation}
\label{eq:quadratic-x-curve}
    x=x(z)=az^2\ , \qquad y=y(z), \qquad \omega_{0,2}=\frac{dz_1dz_2}{(z_1-z_2)^2}\ ,
\end{equation}
where $a$ is some constant which we can leave unfixed and is ultimately a convention which fix later. A critical fact about this particular class of spectral curves is that they have exactly one ramification point at $z=0$. Using the spectral curve as a starting point we can postulate the existence of a density of states which will be fed into the ODE formalism. In particular, the map between the curve data and the density of states that will suffice for our purposes will be of the form\footnote{This procedure is essentially the same as what has been done in recent works~\cite{Johnson:2026jls,Johnson:2026jgs}, which describes how to take a density of states and write in the form of a spectral curve using the relation $z^2=-E$. In those works it is assumed that $x(z)\propto z^2$ which has a single ramification point also related to the fact that double scaled matrix models contain a single branch point at the edge of the spectrum. This is in contrast to spectral curves such as the one written by Okuyama in~\cite{Okuyama:2023discrete} which has two ramification points related to the two edges of the density of states of the DSSYK model (for such cases we expect this prescription to fail).} :  
\begin{equation}
\label{eq:spectral-curve-to-density}
    \rho_0(E)=\frac{\eta}{\pi\hbar}y(z(x))\vert_{x=\gamma E}=\frac{\eta}{\pi\hbar}
    y\!\left(\sqrt{\frac{\gamma E}{a}}\right)\ .
\end{equation}
Under the assumption that $\rho_0(E)$ has an expansion around $E=0$ of the form given in Eq. (\ref{eq:canonical-density}), we expect an ODE formulation of the problem to exist. The parameters $\eta$ and $\gamma$ are free parameters which will be uniquely fixed by requiring $W_{1,1}(z)dz=\omega_{1,1}(z)$. Once the free parameters are fixed using the $\omega_{1,1}$ base case we say the ODE method reproduces the results of TR if, for any stable choice of $(g,n)$, the following equality holds:  
\begin{equation}
\label{eq:ODE-TR-matching}
    W_{g,n}(z_1,\ldots,z_n)dz_1\cdots dz_n
    =\omega_{g,n}(z_1,\ldots,z_n)\ .
\end{equation}
In particular, it is important to note that the map between the curve and a conjecture density assumes a specific form for the relation between $E$  and $z$ which is given as $az^2=\gamma E$. For any given example, if we proceed with our matching procedure and happen to find that $\gamma=-a$ then this will reproduce the the relation $z^2=-E$ which is a standard uniformization relation used in the literature. Furthermore, the value of $\eta$ can be fixed in accordance to some convention which sets the scaling between the spectral curve and the density of states. A common normalization which is discussed in the context of topological recursion for matrix models in is $y(z)=2\pi i\hbar \rho(E=-z^2)$~\cite{Eynard:2014zxa}.

\section{Reproducing q-deformed WP volumes using the ODE method}
\label{sec:top-first-volumes}
Here we will go over some of the details of the computation to obtain q-deformed from the ODE method using the density given in Eq. (\ref{qDefVolDOS}). The first step is to identify the the set of $t_k$ which is done by expanding the density in Eq. (\ref{qDefVolDOS}) around $E=0$ and writing it in the canonical form given in Eq. (\ref{eq:canonical-density}) and reading off the $C_k$. Then we can translate the expression for $C_k$ to $t_k$ using the relation provided in Eq. (\ref{simpletkCkrelation}), below we list the first few:
\begin{equation}
\begin{split}
    &t_1=\frac{1}{2}\ , \qquad t_2=\frac{3\zeta_q(2)}{2}\ , \qquad t_3=\frac{5}{2}\left(\zeta_q(2)^2-\zeta_q(4)\right)\ ,\\
    &t_4=\frac{35}{12}\left[\zeta_q(2)^3-3\zeta_q(2)\zeta_q(4)+2\zeta_q(6)\right]\ , \\ &t_5=\frac{21}{8}\left[\zeta_q(2)^4-6\zeta_q(2)^2\zeta_q(4)+3\zeta_q(4)^2+8\zeta_q(2)\zeta_q(6)-6\zeta_q(8)\right]\ .\\
\end{split}
\end{equation}
These can then be substituted in Eq. (\ref{5u0primetExpression}) to obtain derivatives of $u_0$ at the Fermi surface $x=\mu=0$. We will have:
\begin{equation}
\label{eq:top-u0-derivatives}
    \begin{split}
        &u_0'(0)=-2\ ,\\
        &u_0''(0)=-24\zeta_q(2)\ ,\\
        &u_0'''(0)=-48\left[13\zeta_q(2)^2+5\zeta_q(4)\right]\ ,\\
        &u_0^{(4)}(0)=-320\left[79\zeta_q(2)^3+69\zeta_q(2)\zeta_q(4)+14\zeta_q(6)\right] \ ,\\
        &u_0^{(5)}(0)=-2880\left[485\zeta_q(2)^4+682\zeta_q(2)^2\zeta_q(4)+79\zeta_q(4)^2+224\zeta_q(2)\zeta_q(6)+42\zeta_q(8)\right] \ .\\
    \end{split}
\end{equation}

These five derivatives are sufficient for all the examples considered in here. The formulas for the $W$'s we compute in our examples are readily found in~\cite{Johnson:2026twg} all we need to do is plug in the derivatives we computed in Eq. (\ref{eq:top-u0-derivatives}).

For example, using the formula for $W_{1,1}(z_1)$ in Eq. (\ref{W11Formula}) we have:
\begin{equation}
    \begin{split}
        &W_{1,1}(z_1)=\frac{1}{8z^4}+\frac{\zeta_q(2)}{2z^2}\ ,\\
    \end{split}
\end{equation}
then using Eq. (\ref{eq:inverse-laplace-volumes}) immediately yields:
\begin{equation}
    V^q_{1,1}(b_1)=\frac{b_1^2+24\zeta_q(2)}{48}\ .
\end{equation}

The sphere with three boundaries is given by:
\begin{equation}
\label{eq:top-W03-ODE}
    W_{0,3}(z_1,z_2,z_3)
    =-\frac{u_0'(0)}{2z_1^2z_2^2z_3^2}
    =\frac{1}{z_1^2z_2^2z_3^2}\ .
\end{equation}
Using Eq. (\ref{eq:inverse-laplace-volumes}) gives
\begin{equation}
\label{eq:top-V03-ODE}
    V^q_{0,3}(b_1,b_2,b_3)=1\ .
\end{equation}

For $W_{1,2}(z_1,z_2)$ we quote the formula in~\cite{Johnson:2026twg}:
\begin{equation}
    \begin{split}
        &W_{1,2}(z_1, z_2) = \frac{1}{z_1^2 z_2^2} \left[ A_{12} + B_{12} \sum_{i=1}^2 \frac{1}{z_i^2} + C_{12} \sum_{i=1}^2 \frac{1}{z_i^4} + D_{12} \frac{1}{z_1^2 z_2^2} \right]\ , \\
&\text{with} \quad A_{12} = \frac{u_0^{(3)}(\mu)}{24 u_0'(\mu)} - \frac{u_0''(\mu)^2}{24 u_0'(\mu)^2} \ , \quad B_{12} = -\frac{u_0''(\mu)}{8} \ , \quad C_{12} = \frac{5}{32} u_0'(\mu)^2 \ , \quad D_{12} = \frac{3}{32} u_0'(\mu)^2\ .\\
    \end{split}
\end{equation}
Plugging in the derivatives we find:
\begin{equation}
    \begin{split}
        &A_{12}=7\zeta_q(2)^2+5\zeta_q(4), \qquad B_{12}=3\zeta_q(2), \qquad C_{12}=\frac{5}{8}, \qquad D_{12}=\frac{3}{8}\ ,\\
    \end{split}
\end{equation}
then using Eq. (\ref{eq:inverse-laplace-volumes}) gives:
\begin{equation}
    V^q_{1,2}(b_1,b_2)=\frac{\left(b_1^2+b_2^2\right)^2}{192}+\frac{b_1^2+b_2^2}{2}\zeta_q(2)+7\zeta_q(2)^2+5\zeta_q(4)
\end{equation}

For $W_{2,1}(z_1)$ we have the formula~\cite{Ahmed:2025lxe,Johnson:2026twg}:
\begin{equation}
    \begin{split}
        &W_{2,1}(z_1)=\sum_{k=1}^5\frac{C_{k}}{z_1^{2k}}\ ,\\
        &C_1= \frac{64 u_0''(\mu)^4 - 111 u_0'(\mu) u_0''(\mu)^2 u_0^{(3)}(\mu) + 21 u_0'(\mu)^2 u_0^{(3)}(\mu)^2 + 31 u_0'(\mu)^2 u_0''(\mu) u_0^{(4)}(\mu) - 5 u_0'(\mu)^3 u_0^{(5)}(\mu)}{2880 u_0'(\mu)^5}\ , \\
        &C_2= \frac{17 u_0''(\mu)^3 - 34 u_0'(\mu) u_0''(\mu) u_0^{(3)}(\mu) + 15 u_0'(\mu)^2 u_0^{(4)}(\mu)}{1920 u_0'(\mu)^3}\ , \\
        &C_3=\frac{3 u_0''(\mu)^2 - 29 u_0'(\mu) u_0^{(3)}(\mu)}{768 u_0'(\mu)}, \qquad C_2= \frac{203 u_0'(\mu) u_0''(\mu)}{1536}, \qquad C_5=-\frac{105u_0'(\mu)^3}{1024}\ .\\
    \end{split}
\end{equation}
Plugging in the derivatives we find:
\begin{equation}
    \begin{split}
        &C_1=\frac{845 \zeta_q(2)^4 + 2394 \zeta_q(2)^2 \zeta_q(4) + 555 \zeta_q(4)^2 + 1624 \zeta_q(2) \zeta_q(6) + 630 \zeta_q(8)}{12}\ ,\\
        &C_2=\frac{191 \zeta_q(2)^3 + 243 \zeta_q(2) \zeta_q(4) + 70 \zeta_q(6)}{4}\ ,\\
        &C_3=\frac{359 \zeta_q(2)^2 + 145 \zeta_q(4)}{16},\qquad C_4=\frac{203 \zeta_q(2)}{32}, \qquad C_5=\frac{105}{128}\ .\\
    \end{split}
\end{equation}
Which will then give:
\begin{equation}
    \begin{split}
        &V^q_{2,1}(b)=\frac{b^8}{442368}+\frac{29\zeta_q(2)b^6}{23040}+\frac{(359\zeta_q(2)^2+145\zeta_q(4))b^4}{1920}\notag\\
    &+\frac{(191\zeta_q(2)^3+243\zeta_q(2)\zeta_q(4)+70\zeta_q(6))b^2}{24}\notag\\
    &+\frac{845\zeta_q(2)^4+2394\zeta_q(2)^2\zeta_q(4)+555\zeta_q(4)^2+1624\zeta_q(2)\zeta_q(6)+630\zeta_q(8)}{12}\ ,\notag\\
    \end{split}
\end{equation}

For $W_{0,4}(z_1,z_2,z_3,z_4)$ the formula is given by~\cite{Johnson:2026twg}:
\begin{equation}
\begin{split}
    &W_{0,4}(z_1,z_2,z_3,z_4)=A_{04}\frac{1}{z_1^2z_2^2z_3^2z_4^2}+B_{04}\frac{\sum_{i=1}^4\prod_{j\neq i}z_j^2}{z_1^4z_2^4z_3^4z_4^4}\ ,\\
    &A_{04}(\mu)=-\frac{u_0''(\mu)}{2}, \qquad B_{04}(\mu)=\frac{3}{4}u_0'(\mu)^2\ .\\
\end{split}
\end{equation}
Using the expressions for the derivatives we have:
\begin{equation}
    A_{04}=12\zeta_q(2), \qquad B_{04}=3\ ,
\end{equation}
which will give:
\begin{equation}
    \begin{split}
        &V^q_{0,4}(b_1,b_2,b_3,b_4)=\frac{1}{2}\sum_{i=1}^4b_i^2+12\zeta_q(2)\ ,\notag\\
    &V^q_{1,3}(b_1,b_2,b_3)=\frac{\sum_{i=1}^3b_i^6}{1152}+\frac{\sum_{i=1}^3\left(b_i^4\sum_{j\neq i}b_j^2\right)}{192}+\frac{b_1^2b_2^2b_3^2}{96}\notag\\
    &+\left[\frac{\sum_{i=1}^3b_i^4+3\sum_{i=1}^3\left(\prod_{j\neq i}b_j^2\right)}{4}\right]\zeta_q(2)+\frac{3\sum_{i=1}^3b_i^2}{2}\left(11\zeta_q(2)^2+5\zeta_4(q)\right)\notag\\
    &+\frac{8}{3}\left(76\zeta_q(2)^3+105\zeta_q(2)\zeta_q(4)+35\zeta_q(6)\right)\ .
    \end{split}
\end{equation}

As a final example we can consider the expression for $W_{1,3}$ the formula is given as~\cite{Johnson:2026twg}:
\begin{equation}
\begin{split}
W_{1,3}(z_1, z_2, z_3) &= \frac{1}{z_1^2 z_2^2 z_3^2} \left[ A_{13}(\mu) + B_{13}(\mu) \sum_{i=1}^3 \frac{1}{z_i^2} + M_{13}(\mu) \left( 10 \sum_{i=1}^3 \frac{1}{z_i^4} + 9 \sum_{1 \leq i < j \leq 3} \frac{1}{z_i^2 z_j^2} \right) \right. \\
&\qquad \left. + N_{13}(\mu) \left( 35 \sum_{i=1}^3 \frac{1}{z_i^6} + 30 \sum_{i \neq j} \frac{1}{z_i^4 z_j^2} + 18 \frac{1}{z_1^2 z_2^2 z_3^2} \right) \right], \\[2ex]
A_{13}(\mu) &= \frac{u_0^{(4)}(\mu)}{24 u_0'(\mu)} - \frac{u_0''(\mu) u_0^{(3)}(\mu)}{8 u_0'(\mu)^2} + \frac{u_0''(\mu)^3}{12 u_0'(\mu)^3} \,, \\[2ex]
B_{13}(\mu) &= -\frac{3}{16} u_0^{(3)}(\mu) + \frac{u_0''(\mu)^2}{16 u_0'(\mu)} \,, \\[2ex]
M_{13}(\mu) &= \frac{1}{16} u_0'(\mu) u_0''(\mu) \,, \qquad N_{13}(\mu) = -\frac{1}{64} u_0'(\mu)^3\ .
\end{split}
\end{equation}
Using the expressions we had for the derivatives we obtain:
\begin{equation}
\begin{split}
    &A_{13}=\frac{8 \left( 76 \zeta_q(2)^3 + 105 \zeta_q(2) \zeta_q(4) + 35 \zeta_q(6) \right)}{3}\ ,\\
    &B_{13}=9 \left( 11 \zeta_q(2)^2 + 5 \zeta_q(4) \right)\ ,\\
    &M_{13}=3\zeta_q(2),\qquad N_{13}=\frac{1}{8}\ ,\\
\end{split}
\end{equation}
which will give the following result:
which will give:
\begin{equation}
    \begin{split}
    &V^q_{1,3}(b_1,b_2,b_3)=\frac{\sum_{i=1}^3b_i^6}{1152}+\frac{\sum_{i=1}^3\left(b_i^4\sum_{j\neq i}b_j^2\right)}{192}+\frac{b_1^2b_2^2b_3^2}{96}\notag\\
    &+\left[\frac{\sum_{i=1}^3b_i^4+3\sum_{i=1}^3\left(\prod_{j\neq i}b_j^2\right)}{4}\right]\zeta_q(2)+\frac{3\sum_{i=1}^3b_i^2}{2}\left(11\zeta_q(2)^2+5\zeta_4(q)\right)\notag\\
    &+\frac{8}{3}\left(76\zeta_q(2)^3+105\zeta_q(2)\zeta_q(4)+35\zeta_q(6)\right)\ .
    \end{split}
\end{equation}

All the example volumes computed using the ODE method with the density in Eq. (\ref{qDefVolDOS}) exactly match with the expressions of the q-deformed volumes given in~\cite{DoNorbury:2025dssyk} and will also agree with the volumes generated via topological recursion with the spectral curve given by Do and Norbury in Eq. (\ref{qdefSpectralCuve}).

\section{Properties of $\xi_q(2m)$ as $q\to 1$}
\label{xiqLimitAppendixProof}
To begin, we recall the expression for $\xi_q(2m)$ which was given in Eq. (\ref{xiDefintion}). We split it into two series involving odd and even summation index as shown below:
\begin{equation}
\begin{split}
    &\xi_q(2m)=\sum_{n=1}^\infty\frac{(-q)^{(2n-1)m}}{\left[1+(-q)^{2n-1}\right]^{2m}}+\sum_{n=1}^\infty\frac{(-q)^{2nm}}{\left[1+(-q)^{2n}\right]^{2m}}\\
    &=(-1)^m\sum_{n=1}^\infty \frac{q^{(2n-1)m}}{\left(1-q^{2n-1}\right)^{2m}}+\sum_{n=1}^\infty \frac{q^{2nm}}{(1+q^{2n})^{2m}}\ .\\
\end{split}
\end{equation}
The first series has a denominator that goes to zero and generates a divergence when $q\to 1$. To understand the leading order divergent behavior we expand the denominator around $q=1$ and find, $1-q^{2n-1}=(1-q)(2n-1)+\mathcal{O}(1-q)^2$. It following the leading divergence goes as:
\begin{equation}
   (-1)^m\sum_{n=1}^\infty \frac{q^{(2n-1)m}}{\left(1-q^{2n-1}\right)^{2m}}\overset{q\to 1}{=}\frac{(-1)^m}{(1-q)^{2m}}\sum_{n=1}^\infty\frac{1}{(2n-1)^{2m}}=\frac{(-1)^m}{(1-q)^{2m}}\left(1-\frac{1}{2^{2m}}\right)\zeta(2m)\ .  
\end{equation}
The second series will also diverge at $q=1$ but is convergent when $q\in (0,1)$. We can analyze the divergent behavior near $q=1$ by using the following identity:
\begin{equation}
    \frac{q^{2nm}}{(1+q^{2n})^{2m}}=\sum_{k=0}^\infty (-1)^k\binom{2m+k-1}{k}q^{2n(k+m)}\ ,
\end{equation}
and writing:
\begin{equation}
\begin{split}
    & \sum_{n=1}^\infty \frac{q^{2nm}}{(1+q^{2n})^{2m}}=\sum_{n=1}^\infty \sum_{k=0}^\infty (-1)^k\binom{2m+k-1}{k}q^{2n(k+m)} \\
    &=\sum_{k=0}^\infty (-1)^k\binom{2m+k-1}{k}\frac{q^{2(k+m)}}{1-q^{2(k+m)}}\ .\\
\end{split}
\end{equation}
Now we can clearly see the leading divergent behavior by expanding near $q=1$ and keeping the leading terms:
\begin{equation}
    \sum_{n=1}^\infty \frac{q^{2nm}}{(1+q^{2n})^{2m}}\overset{q\to 1}{=}\frac{1}{1-q}\sum_{k=0}^\infty\frac{\left(-1\right)^k}{2(k+m)}\binom{2m+k-1}{k}=\frac{(m-1)!m!}{2(2m)!}\frac{1}{1-q}\ .
\end{equation}
The leading divergent behavior goes as $(1-q)^{-1}$.
Based on the analysis of the two component series we can immediately conclude the following limit as $q\to 1$ (the second series drops out in the $q\to 1$ limit after multiplying by $(1-q)^{2m}$):
\begin{equation}
    \lim_{q\to 1}(1-q)^{2m}\xi_q(2m)=(-1)^m\left(1-\frac{1}{2^{2m}}\right)\zeta(2m)\ .
\end{equation}

\section{Do and Norbury's $\mathcal{N}=1$ q-deformed WP volumes}
\label{DoNorburyqSWPVolComputations}
Here we review and explicitly compute the some of the $\mathcal{N}=1$ q-deformed super Weil--Petersson volumes using formulas from~\cite{DoNorbury:2025qMirzakhani}. We begin by writing out the basic ingredients set out by Do and Norbury. In particular, we have the recursion relation (here we explicitly assume that $s=0$ so we have pure Neveu–Schwarz (NS) boundaries without Ramond marked points):
\begin{equation}
    \begin{split}
        &L_1\hat{V}_{g,n}(L_1,\vec{L}_K)=\sum_{j=2}^n\int_0^\infty x\hat{R}_q(L_1,L_j,x)\hat{V}_{g,n-1}(x,\vec{L}_{K\setminus \{j\}})dx \\
        &+\frac{1}{2}\int_0^\infty\int_0^\infty xy \hat{D}_q(L_1,x,y)\left[\hat{V}_{g-1,n+1}(x,y,\vec{L}_K)+\sum_{\substack{g_1+g_2=g\\I\sqcup J=K}}\hat{V}_{g_1,|I|+1}(x,\vec{L}_I)\hat{V}_{g_2,|J|+1}(y,\vec{L}_{J})\right]dxdy\ ,\\
    \end{split}
\end{equation}
where:
\begin{equation}
    \begin{split}
        &\hat{D}_q(x,y,z)=\hat{H}_q(y+z,x)\ ,\\
        &\hat{R}_q(x,y,z)=\frac{1}{2}\left[\hat{H}_q(z,x+y)+\hat{H}_q(z,x-y)\right]\ ,\\
        &\hat{H}_q(x,y)=\frac{1}{8\psi(q)^2}\sum_{m\space \text{odd}}\left(-1\right)^{(m-1)/2}q^{(m^2-1)/4}\left(q^{m/2}+q^{-m/2}\right)\left(e^{\frac{1}{4}(x-y)(q^{m/2}-q^{-m/2})}-e^{\frac{1}{4}(x+y)\left(q^{m/2}-q^{-m/2}\right)}\right)\ ,\\
        &\psi(q)=\prod_{m=1}^\infty \frac{1-q^{2m}}{1-q^{2m-1}}\ .\\
    \end{split}
\end{equation}
The initial conditions for $s=0$ sector (i.e. pure Neveu–Schwarz (NS) boundaries) for the recursion relation is:
\begin{equation}
\begin{split}
    &\hat{V}_{0,1}(L)=0\ ,\\
    &\hat{V}_{1,1}(L)=1/8\ .\\
\end{split}
\end{equation}
To explicitly use the recursion relation we make use of the following identities that were provided in~\cite{DoNorbury:2025qMirzakhani}:
\begin{equation}
    \begin{split}
        &\int_0^\infty x^{2k+1}\hat{H}_q(x,y)dx=(2k+1)!\sum_{n=0}^k\hat{b}_n\frac{y^{2k+1-2n}}{(2k+1-2n)!}\ ,\\
    \end{split}
\end{equation}
where $\hat{b}_n$ are defined by:
\begin{equation}
    \exp\left[\sum_{m=1}^\infty \frac{\zeta_q^{odd}(2m)}{m}\left(4z^2\right)^m\right]=\sum_{n=0}^\infty \hat{b}_nz^{2n}\ .
\end{equation}
For example, the first three $\hat{b}_n$'s are:
\begin{equation}
    \begin{split}
        &\hat{b}_0=1\ , \qquad \hat{b}_1=4\zeta_q^{odd}(2)\ , \qquad \hat{b}_2=8\left(\zeta_q^{odd}(2)^2+\zeta_q^{odd}(4)\right)\ .\\
    \end{split}
\end{equation}
Now that we have reviewed the essentially formulas we can explicitly apply them and compute a few of Do and Norbury's volumes. Starting with $\hat{V}_{0,3}$ we have:
\begin{equation}
    \begin{split}
        &\hat{V}_{0,3}(L_1,L_2,L_3)=0\ .\\
    \end{split}
\end{equation}
This is because the terms that are involved that appear in the integrals are simply zero. The next one is $\hat{V}_{1,2}$ which is given by doing the following integrals:
\begin{equation}
    \begin{split}
        &L_1\hat{V}_{1,2}(L_1,L_2)=\int_0^\infty x\hat{R}_q(L_1,L_2,x)\hat{V}_{1,1}(x)dx\\
        &=\frac{1}{16}\int_0^\infty x\left[\hat{H}_q(x,L_1+L_2)+\hat{H}_q(x,L_1-L_2)\right]=\frac{\hat{b}_0}{16}\left[L_1+L_2+L_1-L_2\right]=\frac{1}{8}L_1\\
        &\Rightarrow \hat{V}_{1,2}(L_1,L_2)=\frac{1}{8}\ .
    \end{split}
\end{equation}
Next we can compute $\hat{V}_{2,1}$:
\begin{equation}
    \begin{split}
        &L_1\hat{V}_{2,1}(L_1)=\frac{1}{2}\int_0^\infty\int_0^\infty xy\hat{D}_q(L_1,x,y)\left[\hat{V}_{1,2}(x,y)+\hat{V}_{1,1}(x)\hat{V}_{1,1}(y)\right]dxdy\\
        &=\frac{9}{128}\int_0^\infty\int_0^\infty xy\hat{H}_q(x+y,L_1)dxdy\\
        &=\frac{9}{1024}\int_0^\infty dx_+\int_{-x_+}^{x_+}dx_-\left(x_+^2-x_-^2\right)\hat{H}_q(x_+,L_1)\\
        &=\frac{3}{256}\int_0^\infty x_+^3 \hat{H}_q(x_+,L_1)dx_+=\frac{3L_1}{256}\left[L_1^2+24\zeta^{odd}_q(2)\right]\\
        &\Rightarrow \hat{V}_{2,1}(L_1)=\frac{3}{256}\left[L_1^2+24\zeta^{odd}_q(2)\right]\ ,\\
    \end{split}
\end{equation}
where in going from the second to third line we used the change of variables $x=\frac{x_++x_-}{2}$ and $y=\frac{x_+-x_-}{2}$. 

The next volume we can compute is $\hat{V}_{0,4}$. This will also be zero for similar reasons we found for $\hat{V}_{0,3}$. 

Next we consider the computation of $\hat{V}_{1,3}$ which requires the computation of:
\begin{equation}
    \begin{split}
        &L_1\hat{V}_{1,3}(L_1,L_2,L_3)=\int_0^\infty x\hat{R}_q(L_1,L_2,x)\hat{V}_{1,2}(x,L_3)dx+\int_0^\infty x\hat{R}_q(L_1,L_3,x)\hat{V}_{1,2}(x,L_2)dx\\
        &=\frac{1}{8}\int_0^\infty x \left[\hat{R}_q(L_1,L_2,x)+\hat{R}_q(L_1,L_3,x)\right]dx\\
        &=\frac{1}{16}\int_0^\infty x\left[\hat{H}_q(x,L_1+L_2)+\hat{H}(x,L_1-L_2)+\hat{H}_q(x,L_1+L_3)+\hat{H}_q(x,L_1-L_3)\right]dx\\
        &=\frac{1}{16}\left[L_1+L_2+L_1-L_2+L_1+L_3+L_1-L_3\right]=\frac{L_1}{4}\\
        &\Rightarrow \hat{V}_{1,3}(L_1,L_2,L_3)=\frac{1}{4}\ .\\
    \end{split}
\end{equation}

For $\hat{V}_{2,2}$ we have:
\begin{equation}
    \begin{split}
        &L_1\hat{V}_{2,2}(L_1,L_2)=\int_0^\infty x\hat{R}_q(L_1,L_2,x)\hat{V}_{2,1}(x)dx\\
    &+\frac{1}{2}\int_0^\infty\int_0^\infty xy\hat{D}_q(L_1,x,y)\left[\hat{V}_{1,3}(x,y,L_2)+\hat{V}_{1,1}(x)\hat{V}_{1,2}(y,L_2)+\hat{V}_{1,1}(y)\hat{V}_{1,2}(x,L_2)\right]dxdy\\
    &=\int_0^\infty x \hat{R}_q(L_1,L_2,x)\left[\frac{3}{256}\left(x^2+24\zeta^{odd}_q(2)\right)\right]dx+\frac{9}{64}\int_0^\infty\int_0^\infty xy\hat{D}_q(L_1,x,y)dxdy\\
    &=\frac{3}{512}\int_0^\infty \left[\hat{H}_q(x,L_1+L_2)+\hat{H}_q(x,L_1-L_2)\right]\left[x^3+24x\zeta^{odd}_q(2)\right]dx\\
    &+\frac{9}{512}\int_0^\infty dx_+\int_{-x_+}^{x_+} dx_-\left(x_+^2-x_-^2\right)\hat{H}_q(x_+,L_1)\\
    &=\frac{9L_1}{256}\left[L_1^2+L_2^2+32\zeta_q^{odd}(2)\right]\\
    &\Rightarrow \hat{V}_{2,2}(L_1,L_2)=\frac{9}{256}\left[L_1^2+L_2^2+32\zeta_q^{odd}(2)\right]\ .\\
    \end{split}
\end{equation}

Finally we arrive at the computation of $\hat{V}_{3,1}$:
\begin{equation}
    \begin{split}
        &L_1\hat{V}_{3,1}(L_1)=\frac{1}{2}\int_0^\infty \int_0^\infty xy\hat{D}_q(L_1,x,y)\left[\hat{V}_{2,2}(x,y)+\hat{V}_{1,1}(x)\hat{V}_{2,1}(y)+\hat{V}_{1,1}(y)\hat{V}_{2,1}(x)\right]dxdy\\
        &=\frac{1}{16}\int_0^\infty dx_+ \int_{-x_+}^{x_+}dx_-(x_+^2-x_-^2)\hat{H}_q(x_+,L_1)\left[\hat{V}_{2,2}(x,y)+\hat{V}_{1,1}(x)\hat{V}_{2,1}(y)+\hat{V}_{1,1}(y)\hat{V}_{2,1}(x)\right]\\
        &=\int_0^\infty dx_+\left(\frac{15}{8192}x_+^5+\frac{51\zeta_q^{odd}(2)}{512}x_+^3\right)\hat{H}_q(x_+,L_1)\\
        &=L_1\left[\frac{15}{8192}L_1^4+\frac{63}{256}\zeta^{odd}_q(2)L_1^2+\frac{9}{128}\left(59\zeta^{odd}_q(2)^2+25\zeta^{odd}_q(4)\right)\right]\\
        &\Rightarrow \hat{V}_{3,1}(L_1)=\frac{15}{8192}L_1^4+\frac{63}{256}\zeta^{odd}_q(2)L_1^2+\frac{9}{128}\left(59\zeta^{odd}_q(2)^2+25\zeta^{odd}_q(4)\right)\ .\\
    \end{split}
\end{equation}
Together these give the results listed in Eq. (\ref{DoNorburyqDefSWPVols}). To conclude this appendix let us verify the volumes we computed satisfy the identity:
\begin{equation}
    \lim_{q\to 1}(1-q)^{2g-2}\hat{V}_{g,n}(b_{1,q},..,b_{n,q})=V_{g,n}^{SWP}(b_1,..,b_n)\ , \qquad b_{i,q}=\frac{b_i}{1-q}\ .
\end{equation}
To evaluate the scaling limit we use the result from Do and Norbury~\cite{DoNorbury:2025qMirzakhani}:
\begin{equation}
    \lim_{q\to 1}(1-q)^{2m}\zeta_q^{odd}(2m)=(1-2^{-2m})\zeta(2m)\ .
\end{equation}
Using it we find:
\begin{equation}
    \begin{split}
        &\lim_{q\to 1}(1-q)^0\hat{V}_{1,1}(b_{1,q})=\frac{1}{8}\ ,\\
        &\lim_{q\to 1} (1-q)^0\hat{V}_{1,2}(b_{1,q},b_{2,q})=\frac{1}{8}\ ,\\
        &\lim_{q\to 1} (1-q)^0\hat{V}_{1,3}(b_{1,q},b_{2,q},b_{3,q})=\frac{1}{4}\ ,\\
         &\lim_{q\to 1} (1-q)^2\hat{V}_{2,1}(b_{1,q})=\frac{3}{256}\left(b_1^2+3\pi^2\right)\ ,\\
         &\lim_{q\to 1} (1-q)^2\hat{V}_{2,2}(b_{1,q},b_{2,q})=\frac{9}{256}\left(b_1^2+b_2^2+4\pi^2\right)\ ,\\
         &\lim_{q\to 1}(1-q)^4\hat{V}_{3,1}(b_{1,q})=\frac{15}{8192}b_1^4+\frac{63\pi^2}{2048}b_1^2+\frac{681\pi^4}{8192}\ .\\
    \end{split}
\end{equation}
The results we get after the scaling limit we will denote as, $V^{SWP}_{g,n}(b_1,..,b_n)$, and these are indeed the $\mathcal{N}=1$ super Weil--Petersson volumes in the pure NS sector\footnote{\label{footnoteN=1Volumeconventions}An astute reader may choose to refer to certain works like~\cite{Fuji:2023wcx} and find that they give slightly different coefficients in volumes than the ones we have here. In particular, to recover the results in the table given in~\cite{Fuji:2023wcx} one can just take the following transformation $V^{SWP}_{g,n}(b_1,..,b_n)\to 2^{2g-2}V^{SWP}_{g,n}(b_1/2,..,b_n/2)$ the transformed volumes will perfectly match those in~\cite{Fuji:2023wcx}. From the perspective of the ODE method, the volumes listed in~\cite{Fuji:2023wcx} can be exactly reproduced up to a $(-1)^n$ convention using the density of states, $\rho_0(E)=\frac{\cosh(2\pi\sqrt{E})}{2\pi\hbar\sqrt{E}}$, the $\mathcal{N}=1$ super Weil--Petersson volumes in Do and Norbury's conventions (which are what we are calling $V_{g,n}^{SWP}$ in this paper) are given by using the density, $\rho_0(E)=\frac{\cosh(\pi\sqrt{E})}{2\pi\hbar\sqrt{E}}$. We can see the difference between the two is just a rescaling of $E$ and $\hbar$ so they are physically coming from the same underlying system.} .

\section{More q-deformed WP volumes from $\mathcal{N}=1$ soft edge}
\label{N=1SoftEdgeVolumes}
Similar to how we obtained the $\mathcal{N}=1$ analogue of q-deformed Weil--Petersson volumes by focusing locally around the hard edge of the $\mathcal{N}=1$ DSSYK model we can also try to apply the same procedure around the soft edge of the density given in Eq. (\ref{N=1DSSYKFullDOS}). This is particularly straightforward and by inspection we obtain the following structure from the soft edge at $E=a$:

\begin{equation}
    \rho_0(\epsilon)=\mathcal{C}\sqrt{\epsilon}\exp\left[-\sum_{m=1}^\infty \frac{\zeta_{-q}(2m)}{m}(\gamma\epsilon)^m\right]
\end{equation}
As usual, we will determine the values of $\mathcal{C}$ and $\gamma$ by analyzing the scaled $q\to 1$ limit. To accomplish this we will need to understand the behavior of $\zeta_{-q}(2m)$ as $q\to 1$. We do this by decomposing into even and odd series as follows:
\begin{equation}
\begin{split}
    &\zeta_{-q}(2m)=\zeta^{odd}_{-q}(2m)+\zeta_{-q}^{even}(2m)\\
    &\zeta_{-q}^{odd}(2m)=(-1)^m\sum_{n=1}^\infty\frac{q^{(2n-1)m}}{(1+q^{2n-1})^{2m}}\\
    &\zeta_{-q}^{even}(2m)=\zeta_{q^2}(2m)\\
\end{split}
\end{equation}
Similar to the analysis we did for $\xi_q$ we can write $\zeta_{-q}^{odd}(2m)$ as follows:
\begin{equation}
\begin{split}
    &\zeta_{-q}^{odd}(2m)=(-1)^m\sum_{n=1}^\infty\sum_{k=0}^\infty (-1)^k\binom{2m+k-1}{k}q^{(2n-1)(k+m)}\\
    &=(-1)^m\sum_{k=0}^\infty (-1)^k\binom{2m+k-1}{k}\frac{q^{k+m}}{1-q^{2(k+m)}}\ .\\
\end{split}
\end{equation}
Therefore the leading divergence as $q\to 1$ will go as $\zeta_{-q}^{odd}(2m)\sim \frac{1}{1-q}$.

On the other hand the leading divergent behavior of $\zeta_{q^2}(2m)$ will be:
\begin{equation}
    \zeta_{q^2}(2m)\overset{q\to 1}{=}\frac{\zeta(2m)}{4^m(1-q)^{2m}}
\end{equation}
Based on this analysis we can see in the following scaled $q\to 1$ limit we have:
\begin{equation}
    \lim_{q\to 1}(1-q)^{2m}\zeta_{-q}(2m)=\lim_{q\to 1}(1-q)^{2m}\zeta_{q^2}(2m)=\frac{\zeta(2m)}{4^m}\ .
\end{equation}
Now we can evaluate the following scaled limit as $q\to 1$ of the density of states:
\begin{equation}
\begin{split}
    &\lim_{q\to 1} \frac{\rho_0((1-q)^2\epsilon)}{1-q}=\mathcal{C}\sqrt{\epsilon}\lim_{q\to 1}\exp\left[-\sum_{m=1}^\infty\frac{(1-q)^{2m}\zeta_{-q}(2m)}{m}\left(\gamma\epsilon\right)^m\right]\\
    &=\frac{2\mathcal{C}}{\pi\sqrt{\gamma}}\sin\left(\frac{\pi\sqrt{\gamma\epsilon}}{2}\right)\ .\\
\end{split}
\end{equation}
We recover the JT density of states when $\gamma=-16$ and $\mathcal{C}=(2\pi\hbar)^{-1}$. Throwing away the irrelevant $\zeta_{-q}^{odd}$ part we can write the following density associated to the soft edge of $\mathcal{N}=1$ DSSYK:
\begin{equation}
    \begin{split}
        &\rho_0(E)=\frac{1}{2\pi\hbar}\sqrt{E}\exp\left[-\sum_{m=1}^\infty \frac{(-16)^m\zeta_{q^2}(2m)}{m}E^m\right]\ .\\
    \end{split}
\end{equation}
Which is structurally similar to the density from the soft edge of DSSYK in Eq. (\ref{qDefVolDOS}). However the volumes associated to this density will now involve $\zeta_{q^2}(2m)$ rather than $\zeta_q(2m)$. Another interesting note to make in relation to Do and Norbury's work in~\cite{DoNorbury:2025qMirzakhani} is how the hard edge of $\mathcal{N}=1$ after throwing irrelevant terms involves the q-deformation series $\zeta_q^{odd}(2m)$. One can observe that for the same $\mathcal{N}=1$ model the soft edge after throwing away irrelevant terms will naturally involve $\zeta_{q^2}(2m)=\zeta_q^{even}(2m)$.

\newpage

\bibliographystyle{JHEP}
\bibliography{q_Deformed_WP_References}

\end{document}